\documentclass[]{interact}

\usepackage{array}
\usepackage{longtable}
\usepackage{needspace}
\usepackage{multirow}
\usepackage{xcolor}
\usepackage[hyphens]{url}
\expandafter\def\expandafter\UrlBreaks\expandafter{\UrlBreaks\do\-}

\usepackage[numbers,sort&compress]{natbib}
\bibpunct[, ]{[}{]}{,}{n}{,}{,}

\makeatletter
\def\NAT@def@citea{\def\@citea{\NAT@separator}}
\makeatother

\begin{document}

\articletype{REVIEW ARTICLE}

\title{From White Canes to Edge AI: A Survey of Assistive Navigation
Technologies for Blind and Low-Vision Individuals}

\author{
\name{Abtin Zandi\textsuperscript{a}, Arian Ashrafi\textsuperscript{b}, Sarvin Nami\textsuperscript{a}, Seyed Sina Negahban\textsuperscript{a}, Amirabbas Entezari\textsuperscript{a}, Parsa Abbasian\textsuperscript{c}, Ariyan Azami\textsuperscript{c}, Parisa Taherzadeh\textsuperscript{c}, Bardia Sabagh Kermani\textsuperscript{a}, Roza Ganjipour\textsuperscript{a} and Hamed Farbeh\textsuperscript{d*}\thanks{*Corresponding Author: Hamed Farbeh. Email: h.farbeh@akhu.uz}}
\affil{\textsuperscript{a}Department of Computer Engineering, Amirkabir University of Technology (Tehran Polytechnic), Tehran, Iran; \textsuperscript{b}University of Toronto, Toronto, Canada; \textsuperscript{c}York University, Toronto, Canada; \textsuperscript{d}Department of Mathematics and Informatics, Al-Khwarizmi University, Urgench, 220100, Uzbekistan}
}


\maketitle

\begin{abstract}
Independent mobility is a foundational requirement for the autonomy of
blind and low-vision (BLV) individuals, yet the assistive technologies
that support it have historically evolved slowly.
This survey systematically reviews 108 peer-reviewed studies published between 2008 and 2026, organizing the field into four
technological eras: (i) a foundational era of passive mechanical aids,
namely the white cane and the guide dog; (ii) a hardware era of
electronic travel aids built on ultrasonic, infrared, and satellite
positioning sensors that alert users to proximity but lack semantic
understanding; (iii) a deep learning era in which convolutional and
transformer-based detectors introduced context-aware scene
understanding, initially tethered to cloud infrastructure; and (iv) a
contemporary edge-native era defined by model compression,
quantization, hardware acceleration, and multi-sensor fusion that
relocate intelligence onto resource-constrained wearable and mobile
devices. For each era we analyze representative systems, sensing
modalities, computational architectures, and reported performance, and
we consolidate the field's model-optimization techniques into a single
technical reference spanning four detector families and twelve
compression strategies.
Across the corpus we find a persistent mismatch between rapid algorithmic progress and slow real-world
adoption, driven by cognitive overload in user interfaces, fragmented
evaluation protocols, and the scarcity of studies involving BLV
participants.
We conclude with open challenges, including seamless
indoor--outdoor transition, standardized benchmarks,
privacy-preserving perception, and the emerging role of multimodal
foundation models, and outline a roadmap for socially acceptable,
user-centered assistive navigation.
\end{abstract}


\begin{keywords}
Assistive technology; blind and low-vision (BLV); obstacle detection; deep learning; edge computing; sensor fusion; accessibility
\end{keywords}

\section{Introduction}
\label{sec:intro}

The pursuit of independent mobility stands as a
foundational requirement for autonomy and daily-living integration
among blind and low-vision (BLV) individuals
\cite{bougheloum2024outdoor,nikanfar2025survey,chavan2026aipowered}.
According to the World Health Organization's \emph{World Report on
Vision}, at least 2.2 billion people live with a near or distance
vision impairment, including approximately 43 million who are blind
and 295 million with moderate-to-severe impairment
\cite{who2019vision,gbd2021blindness}; the associated annual global
productivity loss is estimated at US\$411 billion
\cite{belli2026navigation}.
A prevalence of this magnitude places BLV mobility among the largest
accessibility challenges in global health, and it fixes the scale at
which any credible solution must eventually operate.
The economic figure makes the same case in different units: against a
loss of that size, even incremental gains in independent travel
justify sustained research investment, which is why affordable and
dependable navigation aids are a priority rather than a convenience.
Despite this prevalence, for most of recorded history only two
mobility solutions achieved widespread adoption: the white cane and
the guide dog. Over the past three decades, assistive technologies
have undergone a profound transformation, evolving from passive
physical instruments into intelligent, AI-driven computational
companions \cite{zeng2023lightweight,chavan2026aipowered}.

Technical progress, however, has not translated into proportional
real-world impact. Belli \emph{et al.}~\cite{belli2026navigation}
report that roughly 60\% of developed systems remain at medium
Technology Readiness Levels (TRL~4--6), meaning that they never advance
beyond laboratory prototypes, and that none of the 75 systems they
reviewed reached TRL~7--9. They attribute this stall primarily to a
co-design deficiency: only 39\% of projects involved BLV individuals
during the initial design phase \cite{belli2026navigation}. A parallel
gap persists in seamless indoor--outdoor tracking transitions
\cite{belli2026navigation}, and modern electronic aids frequently
overload the very sensory channels, hearing above all, that BLV
travelers depend on for natural situational awareness
\cite{chavan2026aipowered,real2019navigation,chanana2017assistive}.
Mapping the technical lineage of these systems therefore serves a
practical purpose: it establishes the conceptual blueprint necessary
to engineer robust, socially acceptable navigation aids rather than
yet another shelf-bound prototype
\cite{chavan2026aipowered,belli2026navigation}.

This survey traces BLV assistive navigation as a four-era genealogy,
from passive mechanical aids through hardware electronic travel aids
and cloud-tethered deep learning to edge-native intelligence, and
reads each era through the deficit it inherited and the constraints
it left unresolved. Its distinctive emphasis is deployment: which
detection models can actually run on the devices BLV travelers carry,
at what accuracy cost, and with what evidence from BLV users
themselves. In detail, the survey makes the following contributions:
\begin{enumerate}
    \item \textbf{A four-era genealogical taxonomy} of BLV assistive
    navigation that organizes 108 studies by their sensing,
    computation, and feedback architectures into mechanical aids,
    hardware electronic travel aids (ETAs), cloud-tethered deep
    learning, and edge-native intelligent systems.
    \item \textbf{A consolidated technical reference for edge
    deployment}, unifying the field's model-compression strategies
    (pruning, quantization, knowledge distillation, attention-based
    light-weighting, FPGA acceleration, compiler co-design, and
    metaheuristic weight tuning) with their reported
    efficiency--accuracy trade-offs. To
    our knowledge, no prior BLV survey treats this dimension in
    comparable depth.
    \item \textbf{A chronological analysis of four detector families}
    (R-CNN, YOLO, SSD, DETR) with an explicit edge-mobility verdict for
    each, linking architectural evolution
    to the latency budgets that assistive safety actually imposes.
    \item \textbf{A cross-corpus comparison of complete systems}
    spanning sensing modality, model,
    platform, feedback interface, reported performance,
    and, critically, whether evaluation involved BLV participants.
    \item \textbf{A user-centered and ethical analysis} that treats
    interface simplicity and cognitive load as first-class evaluation
    criteria, explaining the enduring dominance of passive aids and
    deriving design constraints for future systems.
    \item \textbf{An open-challenges roadmap} covering benchmarks,
    indoor--outdoor continuity, privacy-preserving perception, and the
    integration of multimodal foundation models.
\end{enumerate}

The remainder of the paper is organized as follows.
Section~\ref{sec:related} positions this survey against prior
reviews, and Section~\ref{sec:method} details the review methodology,
including the research questions that guide it.
Sections~\ref{sec:era1}--\ref{sec:era4} analyze the four eras.
Section~\ref{sec:systems} compares complete systems,
Section~\ref{sec:datasets} examines datasets and benchmarks,
Section~\ref{sec:users} discusses user-centered evaluation, and
Section~\ref{sec:ethics} addresses ethics and equity.
Section~\ref{sec:challenges} outlines open challenges and future
directions, and Section~\ref{sec:conclusion} concludes the survey.

\section{Related Surveys and Scope}
\label{sec:related}

Several recent surveys cover portions of this space.
Chanana \emph{et al.}~\cite{chanana2017assistive} systematically
reviewed electronic travel aids, distinguishing immediate obstacle
detection from long-range navigation and identifying cognitive load,
cost, and non-user-centric design as dominant failure modes. Real and
Araujo~\cite{real2019navigation} traced navigation systems from early
sonar aids to smartphone- and SLAM-based platforms, emphasizing the
human--machine interface as the field's central bottleneck.
Nikanfar \emph{et al.}~\cite{nikanfar2025survey} surveyed recent
AI-driven wearables, highlighting cost and lighting robustness as
persistent limitations. Belli \emph{et al.}~\cite{belli2026navigation}
evaluated 75 studies and nine commercial products, quantifying the
prototype-to-product gap, while Chavan \emph{et
al.}~\cite{chavan2026aipowered} analyzed the trajectory toward
conversational, multi-modal companions. Alarood and
Ibrahim~\cite{alarood2026factors} approached the problem from the
criteria side, ranking safety, responsiveness, reliability, and
usability as the determinants of effectiveness. Within this journal, Bhandari \emph{et al.}~\cite{bhandari2021object} compared deep-learning systems used with navigational tools and compiled a taxonomy of their features, Qin \emph{et al.}~\cite{qin2026wayfinding} conducted a scoping review of how wayfinding and navigation approaches were empirically evaluated in 50 studies, and Panwar \emph{et al.}~\cite{panwar2026understanding} reviewed the functional features of AI-powered assistive technologies across 110 studies published between 2017 and 2025. Bernini \emph{et
al.}~\cite{bernini2014realtime} surveyed stereo-vision obstacle
detection in the adjacent autonomous-vehicle domain, and earlier
typological work catalogued the ETA design space before deep learning
\cite{roentgen2008inventory,hersh2008assistive}.

As Table~\ref{tab:surveys} summarizes, the closest precedent is Real
and Araujo~\cite{real2019navigation}, who trace the field
historically but predate the edge-deployment literature entirely. No
prior review combines a continuous technological genealogy, in
which each era is defined by the deficit it inherited, with depth on
the edge-optimization techniques that now determine whether
assistive perception can run in real time on the devices BLV users
actually carry, such as pruning, quantization, distillation, and
hardware co-design. Those two dimensions define the scope and contribution
of the present survey; for two further dimensions, sensor fusion and
original user-study synthesis, we mark our own coverage as partial,
since the former is treated in a single section without systematic
coding and the latter rests on an audit of the 22 systems in
Table~\ref{tab:systems} rather than the full corpus.

\begin{table}[!t]
\tbl{Comparison of this survey with prior surveys of BLV assistive
navigation. \checkmark~= covered in depth; $\sim$~= partial; $\times$~= not
covered. Coverage was assessed by the present authors from each
survey's own stated scope and contents, and the final row is therefore
a self-assessment applying the same criteria; we mark our own partial
coverage where it exists rather than claiming uniform depth.}
{\setlength{\tabcolsep}{4pt}
\begin{tabular}{lcccccccc}
\toprule
Survey & Year & Historical & Hardware & Deep & Edge & Sensor & User & Ethics/\\
       &      & genealogy  & ETAs     & learning & optimization & fusion & studies & privacy\\
\midrule
Roentgen \emph{et al.} \cite{roentgen2008inventory} & 2008 & $\sim$ & \checkmark & $\times$ & $\times$ & $\times$ & $\sim$ & $\times$\\
\midrule
Bernini \emph{et al.} \cite{bernini2014realtime} & 2014 & $\times$ & $\sim$ & $\times$ & $\times$ & $\sim$ & $\times$ & $\times$\\
\midrule
Chanana \emph{et al.} \cite{chanana2017assistive} & 2017 & $\sim$ & \checkmark & $\times$ & $\times$ & $\sim$ & \checkmark & $\sim$\\
\midrule
Real \& Araujo \cite{real2019navigation} & 2019 & \checkmark & \checkmark & $\sim$ & $\times$ & $\sim$ & \checkmark & $\sim$\\
\midrule
Bhandari \emph{et al.} \cite{bhandari2021object} & 2021 & $\times$ & $\times$ & \checkmark & $\times$ & $\times$ & $\times$ & $\times$\\
\midrule
Nikanfar \emph{et al.} \cite{nikanfar2025survey} & 2025 & $\sim$ & $\sim$ & \checkmark & $\times$ & $\sim$ & $\sim$ & $\sim$\\
\midrule
Belli \emph{et al.} \cite{belli2026navigation} & 2026 & $\sim$ & \checkmark & \checkmark & $\times$ & \checkmark & \checkmark & $\sim$\\
\midrule
Chavan \emph{et al.} \cite{chavan2026aipowered} & 2026 & $\sim$ & $\sim$ & \checkmark & $\sim$ & \checkmark & $\sim$ & \checkmark\\
\midrule
Alarood \& Ibrahim \cite{alarood2026factors} & 2026 & $\times$ & $\sim$ & $\sim$ & $\times$ & $\sim$ & \checkmark & $\sim$\\
\midrule
Qin \emph{et al.} \cite{qin2026wayfinding} & 2026 & $\times$ & $\sim$ & $\times$ & $\times$ & $\times$ & \checkmark & $\times$\\
\midrule
Panwar \emph{et al.} \cite{panwar2026understanding} & 2026 & $\times$ & \checkmark & \checkmark & $\sim$ & $\sim$ & $\sim$ & $\sim$\\
\midrule
\textbf{This survey} & 2026 & \checkmark & \checkmark & \checkmark & \checkmark & $\sim$ & $\sim$ & \checkmark\\
\bottomrule
\end{tabular}}
\label{tab:surveys}
\end{table}

\section{Materials and methods}
\label{sec:method}

We follow the Kitchenham guidelines for systematic literature reviews
in computer science \cite{kitchenham2007guidelines} and report study
selection using the PRISMA 2020 statement \cite{page2021prisma}.

\subsection{Research Questions}
The review is guided by four research questions:
\begin{itemize}
    \item \textbf{RQ1:} How have sensing and computational
    architectures for BLV navigation evolved across technological
    generations, and what capability did each generation add or fail
    to add?
    \item \textbf{RQ2:} Which model-optimization and edge-deployment
    strategies enable real-time on-device inference on
    resource-constrained wearable and mobile platforms, and at what
    accuracy cost?
    \item \textbf{RQ3:} To what extent are proposed systems validated
    with BLV users, on which datasets, and under which evaluation
    protocols?
    \item \textbf{RQ4:} What sociotechnical barriers, such as cognitive load,
    cost, privacy, and infrastructure dependence, explain the gap between
    laboratory performance and real-world adoption?
\end{itemize}

\subsection{Terminology}
We use the abbreviation BLV (blind and low-vision) throughout and mix
person-first and identity-first constructions, in line with current
guidance in the accessibility literature, which recognizes both as
acceptable and defers to the expressed preferences of self-advocacy
groups. Comparison groups are described as ``sighted'' or
``non-disabled'' rather than ``normal'' or ``healthy.''

\subsection{Search Strategy}
Eligibility was defined using the Population--Concept--Context (PCC)
framework: the \emph{population} comprises individuals with any degree
of visual impairment; the \emph{concept} comprises computational
perception and navigation methods (object detection, scene
recognition, localization, sensor fusion, model compression); and the
\emph{context} comprises daily-living spatial navigation in indoor and
outdoor environments. A pilot search on Google Scholar refined the
indexing vocabulary, after which full searches were executed on IEEE
Xplore and the ACM Digital Library between October 2025 and July 2026
using the following query:
\texttt{("visually impaired" OR "blind" OR "BLV") AND ("obstacle
detection" OR "navigation" OR "electronic travel aid") AND ("deep
learning" OR "YOLO" OR "sensor fusion")}.

\subsection{Inclusion and Exclusion Criteria}
We included peer-reviewed journal articles and conference proceedings
published between 2008 and 2026 that propose, evaluate,
or review computational assistive navigation for BLV users. Grey
literature (government documents, newsletters, and non-peer-reviewed
whitepapers) was excluded. Two categories of adjacent-domain work
were retained by explicit protocol decision: (i) lightweight-detection
studies from agriculture and autonomous driving
\cite{yang2025mfd,zeng2023lightweight,yu2024lpyolo,li2024improved,ma2024apple,zhou2022mobileyolo,ju2025lsyolo,huang2025lightyolo,vaithianathan2024fpga},
because their compression and attention techniques transfer directly
to the resource envelopes of assistive edge devices; and (ii)
mobile-systems work on real-time on-device recognition
\cite{chen2015glimpse,cai2021yolobile,yi2013detection,primbetov2024logo,reis2023flying},
which supplies the deployment substrate that assistive systems
inherit. These
inclusions serve RQ2 and are analyzed only for their transferable
techniques, not as assistive systems in their own right.

\subsection{Screening and Extraction}
\begin{figure}[!t]
\centering
\includegraphics[width=0.6\columnwidth,keepaspectratio]{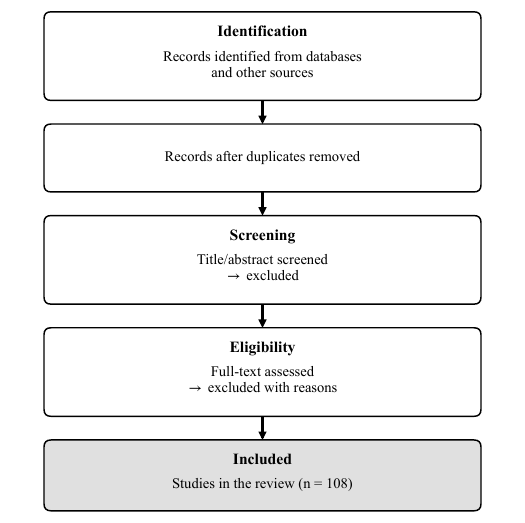}
\caption{PRISMA flow diagram of the study selection process
\cite{page2021prisma}, from initial identification through to the 108
studies synthesized in this survey.}
\label{fig:prisma}
\end{figure}

\begin{figure}[!t]
\centering
\includegraphics[width=\textwidth,keepaspectratio]{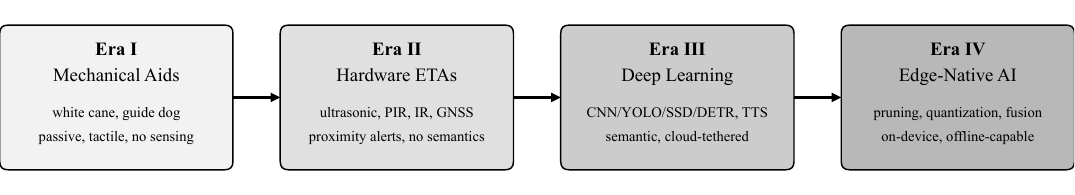}
\caption{The four-era genealogy of BLV assistive navigation
technologies used to organize this survey. Each era is defined by its
dominant sensing substrate, computational paradigm, and connectivity
assumption; representative systems for each era are compared in
Table~\ref{tab:systems}.}
\label{fig:taxonomy}
\end{figure}

The final corpus contains 108 studies. Screening is documented in the
PRISMA flow diagram (Fig.~\ref{fig:prisma}). From each study we
extracted sensing modalities, computational model, deployment
platform, dataset, reported metrics, feedback interface, and the
presence of BLV user evaluation, using a shared extraction sheet.
Following the analyst--reviewer protocol adopted by Belli \emph{et
al.}~\cite{belli2026navigation}, each record was coded by one author
and independently cross-checked by a second, with disagreements resolved by consensus.
The included studies are organized into the four eras shown in
Fig.~\ref{fig:taxonomy}.

\section{Results}

\subsection{Era I: The Foundational Era of Mechanical Aids}
\label{sec:era1}

For most of recorded history, only two mobility solutions achieved
broad adoption among BLV travelers: the white cane and the guide dog.
Understanding \emph{why} these
passive tools endure is not historical trivia; it yields the design
constraints against which every subsequent technology in this survey
should be judged.

\subsubsection{The White Cane}
The white cane achieved widespread adoption through simplicity,
operational reliability, and low cost
\cite{ikram2025obstacle,manduchi2018mobility}. As a passive mechanical
implement, it extends the user's tactile awareness, providing
instantaneous haptic feedback whenever it meets an obstacle, an indentation, or a
change in elevation.
By translating physical contact into raw tactile signals, the cane
informs the traveler of changes in ground texture, variations in
surface level, and immediate structural barriers
\cite{manduchi2018mobility}. This reliability is so robust that early
electronic travel aids and satellite positioning systems were
explicitly conceptualized as high-level navigational
\emph{complements} to the cane rather than replacements: GNSS and GPS
route guidance provide no obstacle detection of their own and cannot
substitute for the immediate physical security of a manual tool
\cite{real2019navigation}.

\subsubsection{The Guide Dog}
Alongside the manual cane, guide dogs represent the primary living
framework for non-visual wayfinding and orientation. These companion
animals offer a dynamic, highly adaptive form of path guidance that
relies on a trained biological partnership rather than the rigid
structural limitations of the cane: actively scanning the path ahead,
identifying open routes, navigating around static and dynamic
obstacles, and stopping reliably at crossings, which allows blind
travelers to achieve fluid locomotion through complex urban settings
\cite{manduchi2018mobility,hersh2008assistive}.

Because both aids remain deeply valid, highly trusted, and widely
used within the global BLV population even at this stage of the deep
learning expansion \cite{belli2026navigation}, any credible account of
the field must explain their persistence rather than treat them as
merely historical.

\subsubsection{Robotic Guide Dogs: An Emerging Hybrid}
A recent line of work replaces the biological partnership with a
quadrupedal robot that reproduces the guide dog's leading behavior
while removing the breeding, training, and care burden
\cite{due2023guidedog,hong2022guidedogrobot}. Due's comparative video
ethnography of blind travelers navigating a street environment with
either a guide dog or a remotely operated robot dog is instructive
about what actually transfers \cite{due2023guidedog}: both agents
support mobility, but they differ in the kind of agency and
sense-ability they contribute, and navigation succeeds as a
distributed, co-constructed achievement between traveler and agent
rather than as delegation to the agent. That finding cuts against the
implicit assumption of much of the corpus surveyed
here, namely that an assistive system should replace the traveler's
perceptual work rather than participate in it, and it anticipates the
interdependence framing discussed in Section~\ref{sec:ethics}.

These platforms sit between Era~I and Era~IV: they inherit
the guide dog's interaction model but depend entirely on the
perception stack of Section~\ref{sec:era4}
\cite{hong2022guidedogrobot}. Belli \emph{et
al.}~\cite{belli2026navigation} note that autonomous robotic
companions account for only 7.3\% of surveyed solutions, constrained
by mechanical cost and transport practicality.

\subsubsection{Why Passive Aids Endure: Simplicity and Cognitive Load}
The enduring dominance of these classic, passive tools over modern
digital alternatives is closely tied to human cognitive processing
limits and user acceptance behavior. Steep learning curves associated
with new digital interfaces discourage long-term adoption,
particularly among older users, who constitute the majority of the BLV
population \cite{chavan2026aipowered,who2019vision}. Rather than
requiring users to decipher algorithmic feedback or configure
technical settings, well-designed aids must respect the preference for
simplicity that many travelers express directly
\cite{chavan2026aipowered}.

The corresponding failure mode of contemporary systems is cognitive
over-saturation. Modern electronic mobility aids frequently emit a
continuous stream of audio notifications, warnings, and localized
tactile pulses; this information overload is a critical challenge in
system design, because excessive environmental information can
saturate a user's remaining sensory channels and measurably
\emph{impair} navigation performance
\cite{chavan2026aipowered,real2019navigation,chanana2017assistive}.
Adoption of software-heavy systems, deep learning frameworks
included, remains correspondingly low despite substantial technical
advances \cite{chavan2026aipowered,belli2026navigation}. Canes and
guide dogs avoid this bottleneck entirely by leaving the user's
hearing unobstructed, preserving access to natural acoustic cues such
as parallel traffic flow and spatial echoes. Hearing functions here as
a second channel of spatial awareness, registering objects outside any
camera's field of view, a point developed further in
Section~\ref{ssec:fusion}. We treat this \emph{simplicity constraint}
as a recurring evaluation criterion throughout the survey.

\subsubsection{Limitations of Mechanical Aids}
The cane's structural shortfalls are most acute in dense,
unpredictable environments. The manual sweeping method creates
distinct challenges in crowded urban spaces, where handling a long
cane safely becomes uncomfortable, and functionally restrictive \cite{chanana2017assistive,manduchi2018mobility}.
Because the cane samples only the ground plane, it leaves the
traveler exposed to overhead and head-level hazards such as hanging
barriers, low signs, and protruding objects positioned directly in the
path \cite{kim2016overhead,chanana2017assistive}. Later systems
explicitly target this gap \cite{tsantikidou2025aal,ikram2025obstacle}.

Guide dogs, in turn, introduce socioeconomic and logistical barriers
that limit global distribution. Specialized breeding programs,
multi-year professional training, lifetime veterinary care, and daily
upkeep create economic barriers that place the option out of reach for
most of the population \cite{hersh2008assistive,belli2026navigation}:
one major guide-dog school reports a cost of approximately
US\$60{,}000 to breed, raise, and train a single dog, an expense
absorbed by the organization rather than the recipient
\cite{gda2026guidedogcost}.

\subsubsection{Synthesis}
Neither aid can read text signs, identify building entrances, name
surrounding landmarks, or recalculate a route around an unexpected
construction barrier, leaving the traveler dependent on memorized
routes or verbal assistance \cite{chavan2026aipowered}. These specific
deficits (overhead detection, crowd navigation, cost, and the
complete absence of semantic environmental information) form the
requirements backlog inherited by the electronic era.

\subsection{Era II: Hardware-Driven Electronic Travel Aids}
\label{sec:era2}

Independent pedestrian mobility is a core requirement for
socioeconomic inclusion, yet navigating unknown spaces without sight
carries high risks of upper-body collision and drop-off accidents
\cite{chanana2017assistive}. To mitigate these hazards, accessibility
researchers engineered a generation of active and passive electronic
travel aids (ETAs)
\cite{chanana2017assistive,real2019navigation,roentgen2008inventory}.
These standalone devices construct a hardware-centric framework that
supplements or substitutes the long cane, extending the traveler's
sensory window through specialized ranging and positioning components
\cite{chanana2017assistive,kasabekar2024smart}. Their defining
characteristic, and their defining limitation, is that they rely on
geometric thresholding rather than learning: they can report that
\emph{something} is 80~cm ahead, but not \emph{what} it is
\cite{kasabekar2024smart,das2020realtime,ikram2025obstacle}.

\subsubsection{Macro-Level Outdoor Path Navigation}
Outdoor navigation across macro-level environments requires tracking
systems capable of resolving geometric coordinates over long ranges
\cite{chanana2017assistive,real2019navigation}. Hardware architectures
in this domain depend fundamentally on the Global Navigation Satellite
System (GNSS): multi-constellation orbital satellites broadcasting
high-frequency microwave signals to determine terrestrial coordinates
\cite{real2019navigation}. Within standalone accessibility
configurations, the Global Positioning System (GPS) functions as the
satellite sub-system integrated directly into ETAs to calculate
absolute latitude and longitude
\cite{kasabekar2024smart,real2019navigation}. First proposed for blind
navigation in 1985 \cite{real2019navigation}, hardware nodes such as
the GPS Neo-6M module capture geospatial telemetry to map a
traveler's progress along an open-air route
\cite{kasabekar2024smart,chintu2026realtime}. High-end systems fuse
satellite records with differential corrections (DGPS) or dead
reckoning, achieving a baseline tracking accuracy of approximately one
meter under optimal conditions
\cite{chanana2017assistive,real2019navigation}.

Kasabekar \emph{et al.}~\cite{kasabekar2024smart} exemplify the mature
form: three ultrasonic sensors covering frontal and lateral directions
(10--300~cm effective range) with distance-dependent haptic
and audio patterns, plus a GPS module and panic button transmitting
location by SMS. Chintu \emph{et al.}~\cite{chintu2026realtime}
generalize the pattern, coupling continuous ultrasonic monitoring with
simultaneous GPS positional data so that instant hazard awareness and
broader spatial orientation are addressed by a single device, with
location tracking additionally enabling external monitoring when
required.

Satellite positioning provides unconstrained coverage across vast
open-air layouts without local infrastructure
\cite{chanana2017assistive,real2019navigation}, but suffers structural
signal loss inside closed architectural spaces, multi-story
buildings, and deep urban canyons due to line-of-sight obstruction
\cite{pham2016realtime,real2019navigation}. It also exhibits inherent
inaccuracy within the final 10--30~m destination window, caused by
atmospheric conditions and multipath reflection
\cite{chanana2017assistive,real2019navigation}.

\subsubsection{Indoor and Infrastructure-Based Local Positioning}
When satellite attenuation prevents reliable macro-level tracking,
ETAs switch to local transceivers to maintain spatial layout data
\cite{pham2016realtime,real2019navigation}. Bluetooth Low Energy (BLE)
beacon systems fix hardware nodes along walls or corridor entries to
transmit proximity notifications to a user's receiver; NavCog
exemplifies this deployment pattern, mapping architectural zones by
detecting when the user enters a beacon's broadcast range
\cite{ahmetovic2016navcog,real2019navigation}.
RFID systems employ passive micro-tags anchored to ceilings, doors, or
floors that respond to electromagnetic interrogation, drawing
operating power from the reader's field to return localized room
indices \cite{chanana2017assistive,real2019navigation}; PERCEPT, for
example, places passive tags at landmarks and destinations and reads
them with a handheld unit paired to a smartphone
\cite{ganz2012percept}. Wi-Fi
positioning triangulates coordinates from Received Signal Strength
Indicator (RSSI) signatures relative to known access points
\cite{chanana2017assistive,real2019navigation}, and smartphone
systems for BLV users fuse such Wi-Fi fingerprints with inertial,
magnetometer, and barometer readings \cite{moder2014indoor}. Where sub-meter
mapping is required, Ultra-Wide Band (UWB) transceivers transmit short
radio pulses and measure Time of Arrival, achieving 15--20~cm indoor
accuracy \cite{real2019navigation}; the SUGAR system couples UWB
positioning with $A^*$ path planning and acoustic guidance, and was
field-tested with a blind participant \cite{martinezsala2015design}. Magnetic field mapping avoids
beacons entirely by measuring directional distortions caused by a
building's structural steel \cite{real2019navigation}; combined with
BLE beacons, magnetic fingerprinting has reached an average error of
2.72~m in a smartphone wayfinding prototype for BLV users
\cite{rathnayake2022hybrid}. Two vision-based
approaches occupy the opposite ends of the infrastructure-cost
spectrum: ARIANNA tracks a smartphone camera along floor-mounted
visual markers spaced roughly 50~cm apart, fusing marker detections
and inertial readings through an Extended Kalman Filter to keep the
user within that corridor tolerance \cite{croce2014arianna}, whereas
Fusco and Coughlan~\cite{fusco2020indoor} localize a smartphone purely
from existing informational signs and inertial dead reckoning,
reaching roughly 1~m accuracy with five blind travelers and
requiring no new infrastructure at all.
Table~\ref{tab:indoor} consolidates these technologies.

The common weakness is infrastructure dependence. These deployments
demand high upfront installation cost and complex maintenance to
mount, configure, and manage large numbers of tags or active beacons,
and they exhibit environmental rigidity: any shift in facility layout
or furniture can invalidate pre-recorded signal fingerprint maps
\cite{chanana2017assistive,real2019navigation,belli2026navigation}.
Belli \emph{et al.}~\cite{belli2026navigation} quantify the
consequence: 44\% of reviewed solutions require a preliminary training
or calibration phase before deployment, concentrated precisely in
camera-based scene parsing, LiDAR profiling, and Wi-Fi fingerprinting.

\begin{table}[!t]
\tbl{Indoor positioning technologies for BLV navigation
(consolidated from \cite{chanana2017assistive,real2019navigation,belli2026navigation}).}
{\setlength{\tabcolsep}{4pt}
\begin{tabular}{>{\raggedright\arraybackslash}p{1.55cm}>{\raggedright\arraybackslash}p{1.75cm}>{\raggedright\arraybackslash}p{1.6cm}>{\raggedright\arraybackslash}p{2.35cm}}
\toprule
Technology & Signal medium & Typical accuracy & Infrastructure / example\\
\midrule
BLE beacons & Low-energy radio & Room / proximity & Wall-mounted transmitters; NavCog \cite{ahmetovic2016navcog}\\
\midrule
RFID tags & EM induction & $<$1.5 m proximity & Floor/ceiling anchored tags; PERCEPT \cite{ganz2012percept}\\
\midrule
Wi-Fi positioning & 2.4/5 GHz radio & Multi-meter & Existing routers (RSSI fingerprinting) \cite{moder2014indoor}\\
\midrule
Ultra-wideband & High-freq.\ pulses & 15--20 cm & Calibrated anchor beacons; SUGAR \cite{martinezsala2015design}\\
\midrule
Magnetic mapping & Geomagnetic field & Room / corridor & Building steel (passive); BLE hybrid \cite{rathnayake2022hybrid}\\
\midrule
Floor markers & Camera + EKF fusion & Sub-meter corridor & Printed markers; ARIANNA \cite{croce2014arianna}\\
\midrule
Sign-based CV & Camera + IMU & $\approx$1 m & No new infrastructure \cite{fusco2020indoor}\\
\bottomrule
\end{tabular}}
\label{tab:indoor}
\end{table}

\subsubsection{Hybrid Positioning: Coupling GNSS, IMU, and BLE}
No single positioning modality spans the indoor--outdoor boundary,
which motivates coupling them. Inertial measurement units bridge
temporary GNSS blind spots by dead reckoning, while BLE beacons
re-anchor the drifting inertial estimate whenever the user passes a
known node \cite{real2019navigation,belli2026navigation}. Belli
\emph{et al.}~\cite{belli2026navigation} report that GNSS serves as
the primary bridge for users transitioning between environments and
that Wi-Fi and BLE are frequently paired to stabilize indoor tracking;
hybrid systems using two or more technologies account for 33\% of the
reviewed literature, against 53\% single-technology designs. Silva and
Wimalaratne~\cite{silva2025navigation} give the fully worked case,
fusing GNSS with a chest-mounted IMU through an error-state extended
Kalman filter; using the inertial unit's on-board digital motion
processor to pre-filter signals reduced mean position error from
3.3~m to 1.74~m over a 110~m course. The formal machinery
of this fusion is treated in Section~\ref{ssec:fusion}.

\subsubsection{Immediate Obstacle Detection and Proximity Sensing}
Immediate safety protection depends on sensors that actively scan the
traveler's walking path \cite{roentgen2008inventory,chanana2017assistive}.
For high-speed ranging, LiDAR emits focused laser pulses and captures
backscattered photons to compute distance; because a single laser path
has a narrow field of view, it resolves the edge boundaries of solid
structures precisely but samples sparsely
\cite{nikanfar2025survey,real2019navigation}.

To broaden the sensing envelope without mechanical scanning,
standalone systems integrate acoustic ultrasound transceivers such as
the HC-SR04 module \cite{kasabekar2024smart,vennila2025smartiot}. These
operate on an active pulse-echo cycle in which an acoustic pulse is
emitted and reflected back upon striking a barrier
\cite{das2020realtime,vennila2025smartiot}. Because the pulse
traverses the sensor--obstacle path twice, the one-way range
$d_{\mathrm{obs}}$ follows from the round-trip time of flight
$t_{\mathrm{delay}}$ as
\begin{equation}
d_{\mathrm{obs}} \;=\; \frac{v_{\mathrm{sound}} \cdot t_{\mathrm{delay}}}{2},
\label{eq:ultrasonic}
\end{equation}
where $v_{\mathrm{sound}}$ is the speed of sound in air, conventionally
taken as $343~\mathrm{m\,s^{-1}}$ at $20\,^{\circ}$C; because this
value varies by roughly $0.6~\mathrm{m\,s^{-1}}$ per degree Celsius,
uncompensated devices accumulate a systematic ranging bias across
temperature. Eq.~\eqref{eq:ultrasonic} handles close-range
identification across a standard zone from 2~cm to 400~cm
\cite{vennila2025smartiot,das2020realtime}.

\begin{figure}[!t]
\centering
\includegraphics[width=\columnwidth,keepaspectratio]{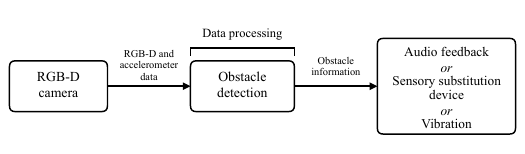}
\caption{Pham \emph{et al.}'s Kinect-based pipeline: an RGB-D camera
feeds obstacle detection via RANSAC floor-plane fitting, with output
routed to audio or a sensory-substitution device. Redrawn after Pham
\emph{et al.}~\cite{pham2016realtime}.}
\label{fig:pham}
\end{figure}

For dense spatial awareness, time-of-flight (ToF) sensors track light
phase delays across semiconductor arrays to extract depth
simultaneously across the frame
\cite{chanana2017assistive,kuriakose2020smartphone}. Fusing these
depth pixels with color yields RGB-D cameras such as the Microsoft
Kinect, which process color indices alongside dense depth matrices in
real time \cite{pham2016realtime}. To isolate the floor plane from
obstacles, Pham \emph{et al.}~\cite{pham2016realtime} apply Random
Sample Consensus (RANSAC), an iterative estimator that repeatedly fits
a candidate model to a minimal random subset of points and retains the
hypothesis commanding the largest consensus set, making it robust to
the outliers that dominate raw depth data. The model fitted here is
the plane
\begin{equation}
ax + by + cz + d = 0,
\label{eq:ransac}
\end{equation}
in which $(a,b,c)$ is the plane normal and $d$ its offset from the
origin; the floor is identified as the consensus plane whose normal is
parallel to the sensor's vertical axis. Points excluded from that
consensus set are then clustered into obstacles, with planes
approximately perpendicular to the floor recovered as walls and doors
(Fig.~\ref{fig:pham}).

Inertial measurement units complement visual tracking when feature
points become occluded or cluttered
\cite{chanana2017assistive,das2020realtime}. IMU chipsets combine
three-axis accelerometers and gyroscopes to capture changes in walking
velocity and body tilt \cite{das2020realtime,chanana2017assistive},
from which dead-reckoning loops compute relative motion increments
\cite{chanana2017assistive,real2019navigation}.

Each modality carries characteristic physical failure modes. Acoustic
wave divergence makes ultrasonic transceivers prone to echo
scattering, producing missed detections on narrow or irregular
surfaces, while optical rangefinders and active infrared emitters
suffer severe performance loss or complete blinding under direct
sunlight \cite{chanana2017assistive,roentgen2008inventory,ikram2025obstacle}.
Conversely, RGB-D and ToF arrays record spatially dense maps that
remain reliable in total darkness or low-light interiors
\cite{pham2016realtime,bai2019wearable,gao2025wearable}.

\subsubsection{Deployed ETA Devices}
Table~\ref{tab:devices} maps the preceding sensing principles onto
the commercial and research devices that implemented them, drawn from
the systematic inventory of Chanana \emph{et
al.}~\cite{chanana2017assistive}. Two design patterns are visible.
First, tactile output dominates: seven of the eight devices convey
distance through vibration rather than speech, preserving the
auditory channel in line with the simplicity constraint of
Section~\ref{sec:era1}. Second, head-level coverage, the deficit
identified in Section~\ref{sec:era1}, is addressed only by the later
entries (UltraCane, Tom Pouce, iGlasses), confirming that the
requirements backlog inherited from the mechanical era drove the
design trajectory of this one.

\begin{table}[!t]
\tbl{Deployed electronic travel aids and the sensing principles
they implement. All rows, including list price at time of review,
consolidated from Chanana \emph{et al.}~\cite{chanana2017assistive};
prices reflect that source and predate this survey, so they indicate
relative affordability rather than current market cost.}
{\begin{tabular}{>{\raggedright\arraybackslash}p{1.31cm}>{\raggedright\arraybackslash}p{2.54cm}>{\raggedright\arraybackslash}p{4.10cm}>{\raggedright\arraybackslash}p{3.03cm}>{\raggedright\arraybackslash}p{1.07cm}}
\toprule
Device & Primary sensor technology & Information conveyed and coverage & User feedback interface & Price\\
\midrule
Miniguide & Ultrasonic transceiver & Distance to the nearest detected obstacle & Continuous vibratory patterns or variable audio & \$399\\
\midrule
PalmSonar & Ultrasonic, asymmetric field & Multi-direction distance to nearest obstacle; asymmetric beam minimizes side-corridor interference (discontinued) & Localized tactile vibration & \$1000\\
\midrule
K Sonar & Multi-tone ultrasonic array & All objects within the active field of view simultaneously; pitch encodes distance & Audio tones of varying pitch and material-specific timbre & \$1085\\
\midrule
Tom Pouce & Multi-transmitter infrared array & Distance to nearest obstacle; shoulder-width paths and head-high protrusions in later versions & Discrete mechanical vibration & \$700--1000\\
\midrule
Teletact & Highly directional laser beam & Precise target distance from 10 cm to 6 m; requires continuous manual scanning & Two vibrators at the first and second control fingers & \$2300\\
\midrule
UltraCane & Dual-transceiver ultrasonic pair & Simultaneous ground-level and head-level proximity & Two independent vibrating buttons on the handle grip & \$825\\
\midrule
SmartCane & Adjustable ultrasonic ranging & Obstacles between knee and head level; adjustable for user height and cane style & Tactile patterns across the whole handle grip & \$60\\
\midrule
iGlasses & Wearable head-level ultrasonic & Distance to obstacles at head and shoulder clearance & Vibration intensity on the user's forehead & \$96\\
\bottomrule
\end{tabular}}
\label{tab:devices}
\end{table}

\subsubsection{Assessment}
Era~II systems satisfied the simplicity constraint, since their alerts
are immediate and interpretable, and they remain attractive for their low cost
\cite{ikram2025obstacle,vennila2025smartiot,chintu2026realtime}. But
they lack semantic awareness entirely: users cannot distinguish a
pedestrian from a pole, or a puddle from a drop-off
\cite{kasabekar2024smart,das2020realtime,ikram2025obstacle,rahman2018obstacle}.
Das \emph{et al.}~\cite{das2020realtime} illustrate the ceiling
precisely: their multisensor cane resolves distance to 400~cm and
infers direction and velocity from PIR arrays, yet cannot identify
what type of obstacle is approaching. Closing that semantic gap
required learning-based vision, the subject of Era~III.

\subsection{Era III: The Deep Learning Revolution}
\label{sec:era3}

\begin{figure}[!t]
\centering
\includegraphics[width=\textwidth,keepaspectratio]{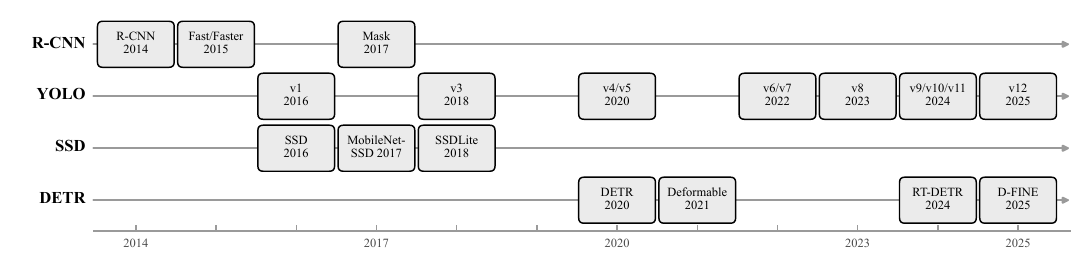}
\caption{Evolution of the four object-detection families used across
the corpus, positioned by publication year (x-axis, 2014--2025).
YOLOv8 (via \cite{wang2024yolood,zhou2022mobileyolo,ikram2025obstacle}
and others), R-CNN (as a comparison baseline
\cite{bougheloum2024outdoor}), SSD/MobileNet-SSD
\cite{suman2022vision,kuriakose2020smartphone,tapu2013smartphone}, and
DETR \cite{ikram2025detr} are the specific versions with direct
assistive-system adoptions documented in this survey; the remaining
nodes mark the family's technical lineage. Edge-mobility verdicts for
each family are summarized in Table~\ref{tab:detectors}.}
\label{fig:lineage}
\end{figure}

Deep learning replaced rule-based geometric processing with adaptive,
context-aware visual pipelines
\cite{bai2019wearable,pham2016realtime,ikram2025detr}. This section
traces the four detector families that made real-time semantic
perception possible (Fig.~\ref{fig:lineage}), reviews their
translation into assistive systems, and examines the era's structural
weakness: cloud dependence.

\subsubsection{Two-Stage Detectors: The R-CNN Family}
The Region-based Convolutional Neural Network (R-CNN) framework was
engineered to bridge the gap between handcrafted feature extraction
(SIFT, HOG) and learned deep representations, recasting detection as a
propose-then-classify workflow \cite{girshick2014rcnn}. The original
2014 architecture generated roughly 2{,}000 class-independent region
proposals per image via selective search, passing each through a CNN
before classification by linear support vector machines
\cite{girshick2014rcnn}. Fast R-CNN (2015) integrated a Region of
Interest pooling layer so the full image traverses the convolutional
extractor only once \cite{girshick2015fast}, and Faster R-CNN (2015)
eliminated selective search entirely by introducing an internal Region
Proposal Network, yielding a unified two-stage framework
\cite{ren2015faster}. Mask R-CNN (2017) added a parallel mask branch
alongside classification and box regression, enabling pixel-level
instance segmentation \cite{he2017mask}.

Despite excellent localization accuracy on standard benchmarks, the
assistive literature documents a decisive migration away from these
multi-stage models. Bougheloum \emph{et
al.}~\cite{bougheloum2024outdoor} report that a single-stage YOLOv5
configuration achieved higher precision, recall, F1, and mAP than an
R-CNN baseline across custom outdoor classes and MS COCO
\cite{lin2014coco}, and that R-CNN struggled to localize multiple
objects simultaneously in cluttered scenes. The dual-stage execution
path incurs memory and latency overheads that make these models
impractical for zero-latency deployment on smartphone chipsets, where
rapid feedback is a safety requirement
\cite{bougheloum2024outdoor,ikram2025detr}.

\subsubsection{Single-Stage Detectors: The YOLO Family}
The You Only Look Once (YOLO) family redefined detection as a unified
single-stage regression task, mapping input pixels directly to
bounding-box coordinates and class probabilities in one forward pass
and thereby removing the region-proposal stage entirely
\cite{redmon2016yolo}. YOLOv3 introduced multi-scale prediction over a
DarkNet-53 backbone, substantially improving small-object recall
\cite{redmon2018yolov3}. YOLOv4 (2020) incorporated Cross-Stage
Partial Networks to accelerate feature aggregation
\cite{bochkovskiy2020yolov4}; YOLOv5 added native PyTorch integration,
automated anchor calculation, and lightweight configurations such as
YOLOv5s \cite{jocher2020yolov5}. YOLOv6 and YOLOv7 (2022) emphasized
hardware-friendly industrial design and structural re-parameterization
\cite{li2022yolov6,wang2023yolov7}, and YOLOv8 (2023) moved the family
to an anchor-free paradigm with decoupled heads
\cite{jocher2023yolov8}.

The recent generation pushes efficiency
further: YOLOv9 integrated Generalized Efficient Layer Aggregation
Networks with Programmable Gradient Information to limit information
loss across depth \cite{wang2024yolov9}; YOLOv10 adopted consistent
dual-label assignment to achieve end-to-end detection free of
non-maximum suppression (NMS) latency \cite{wang2024yolov10}; YOLOv11
introduced C3K2 extraction blocks, optimized SPPF layers, and
convolutional parallel spatial attention to preserve small-target
features in dense scenes \cite{khanam2024yolov11}; and YOLOv12 (2025)
shifted to an attention-centric design combining localized area
attention with residual efficient layer aggregation
\cite{tian2025yolov12}.

Within assistive vision, YOLO variants are deployed most heavily
against small and occluded targets. YOLO-OD integrates a Feature
Weighting Block and an Adaptive Bottleneck Block to differentiate
obstacles of varying size in complex outdoor scenes
\cite{wang2024yolood}; DC-YOLOv8 replaces baseline components with
parallel cascades of standard and depthwise separable convolutions to
capture richer context for small objects \cite{lou2023dcyolov8}; and
MBAB-YOLO introduces a multi-branching blended attention block that
adapts receptive field size to target scale \cite{zhang2023mbab}.
Environmental robustness has been pursued along the same lines:
YOLO-Extreme employs multi-dimensional collaborative attention to
suppress atmospheric fog and emphasize low-contrast obstacles
\cite{wang2025yoloextreme}; LS-YOLO
instead combines multi-scale attention with an adaptive
Wasserstein-based loss to handle fog, rain, snow, and dense traffic
\cite{ju2025lsyolo}; and Light-YOLO targets small obstacles
specifically under low-light nighttime conditions with a Wise-IoU loss
and a lightweight backbone \cite{huang2025lightyolo}.

Two smartphone-only systems illustrate the family's practical
envelope. Obs-tackle runs a customized YOLOv3-tiny network on a
commodity phone, mapping filtered bounding boxes into a geometric grid
that divides the forward path into safety sectors and issuing
progressive zone-aware spoken alerts; it reports
92.3\%~mAP, 93.1\% precision, and
92.6\% recall under daytime outdoor conditions, with
degradation under low light, rain, and motion blur
\cite{vijetha2024obstackle}. Baskar \emph{et
al.}~\cite{baskar2026enhancing} target the ego-motion problem created
by the user's walking gait, pairing YOLOv8n with adaptive multi-cue
weighting over dense optical flow, geometric size-change rates, and
structural similarity; the system reaches
79.9\%~mAP@0.5 at 21~FPS, but profiling shows
global ego-motion estimation consuming 85.3\% of pipeline
time. This is a concrete illustration that in assistive deployment
the bottleneck is often not the detector.

\subsubsection{The SSD Family}
The Single Shot MultiBox Detector was introduced as a one-stage
alternative balancing early YOLO's speed against the localization
accuracy of two-stage architectures, specifically addressing the
inaccurate detection of small or densely grouped objects
\cite{liu2016ssd}. SSD300 and SSD512 predict classes and box offsets
independently across multi-scale feature maps drawn from intermediate
convolutional layers, capturing large entities in deeper layers and
small targets in shallower, higher-resolution ones
\cite{liu2016ssd}. MobileNet-SSD replaced the heavy VGG-16 backbone
with an efficient mobile framework \cite{howard2017mobilenets}, and
SSDLite (2018) substituted depthwise separable convolutions throughout
the prediction head, sharply reducing FLOPs and storage footprint for
edge deployment \cite{sandler2018mobilenetv2}.

In assistive frameworks, the family is used to build stable,
low-latency classification streams directly on mobile and embedded
platforms \cite{kuriakose2020smartphone,suman2022vision}. One
clarification is warranted: although some summaries group Tapu \emph{et
al.}~\cite{tapu2013smartphone} with deep learning configurations, that
system in fact implements HOG descriptors with a bag-of-visual-words
retrieval pipeline and an SVM classifier, not an SSD network; it
nonetheless establishes the latency framing for the family, recording
an average 102~ms per frame on 2013-era smartphone hardware
\cite{tapu2013smartphone}. Kuriakose \emph{et
al.}~\cite{kuriakose2020smartphone} evaluate paired MobileNet+SSDv2
pipelines running locally to eliminate network dependence, and Suman
\emph{et al.}~\cite{suman2022vision} deploy an SSD instance coupled to
a recurrent network inside a smart-cane infrastructure, reporting
95.06\% indoor and 87.68\% outdoor classification
accuracy, although at an execution latency of 4.82~s that
falls well outside any reasonable real-time budget. Leong \emph{et
al.}~\cite{leong2026tensorflow} demonstrate the family's continued
viability at the low end, converting a MobileNetV2-SSD framework to
TensorFlow Lite at $300 \times 300$ input resolution on a Raspberry~Pi
4B, covering eight municipal obstacle classes at
92.5\% accuracy and 85~ms per frame.

\subsubsection{Transformer Detectors: The DETR Family}
The Detection Transformer family was driven by the goal of eliminating
hand-crafted components (anchor boxes, assignment heuristics, and NMS
post-processing) by reframing detection as direct set prediction
\cite{carion2020detr}. The 2020 baseline couples a CNN backbone for
local features with a transformer encoder--decoder in which learnable
object queries interact with global scene context through self- and
cross-attention, optimized by bipartite matching loss; its principal
weakness was slow training convergence \cite{carion2020detr}.
Deformable DETR (2021) resolved this with a deformable attention
module attending only to a sparse set of sampling points around each
target, accelerating convergence by roughly an order of magnitude
\cite{zhu2021deformable}. RT-DETR (2024) achieved the first genuinely
real-time detection transformer through a hybrid encoder with
decoupled intra-scale interaction and cross-scale fusion, bypassing
the NMS bottleneck and matching or exceeding dominant YOLO variants in
both speed and accuracy \cite{zhao2024rtdetr}. D-FINE re-engineered
the regression head with Fine-grained Distribution Refinement,
iteratively refining probability distributions rather than predicting
fixed coordinates, which improves edge localization on unstructured
targets \cite{peng2025dfine}.

Ikram \emph{et al.}~\cite{ikram2025detr} provide the assistive
application: a DETR framework paired with a mini camera and mobile
application for real-time obstacle detection without anchor boxes.
Benchmarked against Faster R-CNN and YOLOv8, the model reports a
confidence score of 99\% and classification precision of
98\% across eighty obstacle categories, and, critically for
this survey's argument, was optimized for edge execution through a
TensorFlow Lite conversion pipeline that stabilized inference at
40~ms per frame \cite{ikram2025detr}.

\subsubsection{Comparative Verdict for Edge Deployment}
Table~\ref{tab:detectors} summarizes the four families against the
constraint that governs this domain: a detection pipeline whose
end-to-end latency exceeds a walking-pace reaction window is unsafe
regardless of its benchmark accuracy. The verdicts synthesize
published evidence rather than a single controlled benchmark; because
that evidence is measured on heterogeneous hardware, the relative
ordering of the four families is more reliable than any absolute
latency figure.

\begin{table}[!t]
\tbl{The four detector families and their suitability for
on-device assistive deployment. Verdicts synthesize the
architectural and empirical evidence in Section~\ref{sec:era3}.
Published latencies are measured on heterogeneous hardware, several
on desktop GPUs rather than mobile silicon; the ordering across
families is the reliable signal, not the absolute figures.}
{\begin{tabular}{>{\raggedright\arraybackslash}p{1.20cm}>{\raggedright\arraybackslash}p{2.10cm}>{\raggedright\arraybackslash}p{3.30cm}>{\raggedright\arraybackslash}p{3.00cm}>{\raggedright\arraybackslash}p{2.40cm}}
\toprule
Family & Key milestones & Architectural principle & Edge-mobility verdict & Assistive uses in corpus\\
\midrule
R-CNN & R-CNN 2014~\cite{girshick2014rcnn}; Fast/Faster 2015~\cite{girshick2015fast,ren2015faster}; Mask 2017~\cite{he2017mask} & Two-stage: region proposal then classification; RPN internalizes proposals & Not recommended on-device; dual-stage path incurs prohibitive memory and latency cost & Used only as a comparison baseline \cite{bougheloum2024outdoor}\\
\midrule
YOLO & v1 2016~\cite{redmon2016yolo} through v12 2025~\cite{tian2025yolov12} & Single-stage unified regression; anchor-free from v8; NMS-free from v10 & Recommended. Deployment evidence in this corpus reaches v8n under int8; the NMS-free v10 and attention-centric v12 nano variants are architecturally well suited to this envelope but are not yet deployed in any reviewed assistive system & Dominant across the corpus \cite{bougheloum2024outdoor,joshi2020efficient,okolo2025smart,wang2024yolood,bastidas2025computer,vijetha2024obstackle,baskar2026enhancing}\\
\midrule
SSD & SSD 2016~\cite{liu2016ssd}; MobileNet-SSD 2017~\cite{howard2017mobilenets}; SSDLite 2018~\cite{sandler2018mobilenetv2} & Single-stage with multi-scale feature-map prediction & Recommended for legacy or budget terminals with constrained memory bandwidth & \cite{kuriakose2020smartphone,suman2022vision,leong2026tensorflow}\\
\midrule
DETR & DETR 2020~\cite{carion2020detr}; Deformable 2021~\cite{zhu2021deformable}; RT-DETR 2024~\cite{zhao2024rtdetr}; D-FINE 2025~\cite{peng2025dfine} & End-to-end set prediction; global attention; no anchors or NMS & Viable in lightweight form (RT-DETR-Nano); baseline models require an NPU or careful quantization & \cite{ikram2025detr}\\
\bottomrule
\end{tabular}}
\label{tab:detectors}
\end{table}

\subsubsection{Assistive Vision Systems}
Translation into deployed systems followed rapidly. Representative
designs pair a wearable or smartphone camera with a detector and a
text-to-speech interface: YOLOv5 with Google TTS trained on MS COCO
plus a custom 10-class outdoor dataset delivering audio to earphones
\cite{bougheloum2024outdoor,lin2014coco}; YOLOv3 with ultrasonic
fallback and mode switching across indoor, outdoor, and text-reading
contexts, reaching 95.19\% accuracy on a custom dataset
\cite{joshi2020efficient}; MobileNet-based recognition with GPS
tracking at 83.3\% accuracy \cite{ashiq2022cnn}; YOLOv8 on a
Raspberry~Pi with GPS voice guidance at 91.7\% average
accuracy \cite{okolo2025smart}; YOLOv5s in a smartphone
application trained on a purpose-built 7{,}600-image obstacle dataset
with combined speech and vibration feedback
\cite{bastidas2025computer}; YOLOv3 with blob detection and
text-to-speech requiring no hardware beyond the phone itself
\cite{kumar2021object}; and a YOLOv5 Android app that extends object
detection to currency-note recognition across eleven denominations
\cite{kumar2023empowering}. Beyond classification, YOLOv8 segmentation
layers have been applied to curb and sidewalk edge detection for
wheelchair and mobility aids \cite{kao2024edge}, and YOLO paired with
multi-object trackers isolates pedestrian behavior across traffic
lanes \cite{mostafi2022jaywalking}.

Scene-level understanding complements object detection. SceneRecog
classifies whole environments (street, corridor, kitchen) using
EfficientNet-Lite, providing high-level context rather than
object-level enumeration \cite{kuriakose2021scenerecog}. Salient-object
detection filters output to navigation-relevant entities only,
directly addressing the information-overload constraint of
Section~\ref{sec:era1}: Mukhiddinov and Cho~\cite{mukhiddinov2021smart}
prepend a two-branch exposure fusion network for low-light
enhancement, then pass only the most meaningful objects to
text-to-speech and tactile output, reducing redundant announcements.
Vision transformers extend semantic reach further
\cite{alarood2025enhanced,ikram2025detr}. Prakash \emph{et
al.}~\cite{prakash2026aipowered} broaden the scope beyond obstacles
entirely, adding a facial recognition module that classifies
individuals as known or unknown against a local database and a GSM
emergency channel transmitting distress coordinates in
2.1--2.6~s, reporting 96\% perception accuracy on
clear paths and 94\% under obstruction.
LLM-Glasses \cite{tokmurziyev2026llmglasses} pairs YOLO-World detection
with GPT-4o scene reasoning and replaces speech with haptic cues on the
glasses' temples, reaching 91.8\%, 84.6\%, and 81.5\% decision accuracy
without, with static, and with dynamic obstacles in controlled trials.

\subsubsection{Deployed Commercial Systems}
Academic surveys of this field routinely omit the products BLV users
actually have installed, which distorts the picture of the state of
the art. Microsoft's Seeing AI \cite{microsoft2025seeingai} and
Google's Lookout \cite{google2018lookout} provide scene description,
text recognition, and object identification with mature accessible
interfaces, blending on-device and cloud processing; BlindSquare
\cite{blindsquare2025} supplies GNSS-based point-of-interest
announcement for pedestrians; and WeWALK \cite{wewalk2025} pairs an
instrumented cane with a companion application, occupying the hybrid
position between Era~I and Era~IV that Section~\ref{sec:era1}
identified. Belli \emph{et al.}~\cite{belli2026navigation} catalog a
further nine devices spanning forehead-vibration cameras, vibrotactile
belts, LiDAR-equipped glasses, and QR-based transit wayfinding.

Two observations follow. First, these products consistently prioritize
interface maturity over detection novelty. This is the opposite of
the research corpus and a direct reflection of the simplicity constraint
of Section~\ref{sec:era1}. Second, their limitations are the ones this
survey attributes to the era: dependence on cloud services, on
device-specific sensors, or on installed infrastructure
\cite{messaoudi2022,belli2026navigation}. That commercial systems and
laboratory prototypes fail along the same axis suggests the constraint
is structural rather than a matter of engineering maturity.

\subsubsection{The Cloud Bottleneck}
Early intelligent systems were heavily cloud-tethered, offloading
inference to remote servers
\cite{bai2017cloud,ikram2025detr,chen2015glimpse}. Bai \emph{et
al.}~\cite{bai2017cloud} exemplify the pattern: a helmet-mounted stereo
camera streaming to a cloud platform running vision-based SLAM,
thousand-class recognition, and OCR, with results returned as speech.
Offloading introduces wireless latency that is not merely inconvenient
but hazardous at walking pace, and renders systems inoperable in
connectivity-poor environments
\cite{bai2017cloud,chen2015glimpse,datta2024enhanced}. Glimpse
\cite{chen2015glimpse} quantified the problem and partially mitigated
it, maintaining an active cache of frames tracked forward from stale
server results; even so, capping in-flight requests to conserve
bandwidth introduced a discovery latency of at least one round trip
whenever a new object entered the scene. Silva and
Wimalaratne~\cite{silva2025navigation} show the same trade-off in a
recent design, offloading label detection to a cloud service at
570--650~ms round trip depending on image compression, an
interval during which a walking user covers roughly half a meter. The
decisive response was to move the models themselves onto the device:
the subject of Era~IV.

\subsection{Era IV: Edge-Native Intelligence}
\label{sec:era4}

The contemporary era relocates computation onto mobile and embedded
hardware
\cite{tsantikidou2025aal,kuriakose2021towards,mukhiddinov2021smart,cai2021yolobile,mramba2026aismartstick},
pursuing three complementary strategies: multi-sensor fusion
(Section~\ref{ssec:fusion}), offline self-sufficiency
(Section~\ref{ssec:offline}), and aggressive model optimization
(Section~\ref{ssec:opt}). Mramba \emph{et
al.}~\cite{mramba2026aismartstick} state the motivating trade-off
plainly: multi-sensor architectures minimize physical blind spots but
raise hardware complexity and power draw without supplying semantic
understanding, whereas embedded deep learning supplies semantics but
has historically depended on cloud topologies that reintroduce
latency. Their resolution, which places a hyper-optimized detector on
local edge hardware alongside a supplementary sensor matrix, is
representative of the era as a whole.

\subsubsection{Multi-Sensor Fusion and SLAM}
\label{ssec:fusion}
Fusion frameworks resolve the blind spots of standalone modalities
\cite{zhang2019radar,mahammad2026fusion,silva2025navigation}. It is
worth distinguishing two architectures that the literature often
conflates: \emph{sensor integration} feeds multiple streams into a
controller as independent inputs serving distinct objectives, whereas
\emph{sensor fusion} combines raw inputs across sensors or across
time into a unified representation that reduces tracking uncertainty.
Silva and Wimalaratne~\cite{silva2025navigation}, adopting this
distinction from the sensor-integration literature, argue that most
assistive navigation work targets integration rather than fusion. Our
reading of the corpus is consistent with that assessment, though we
did not code the distinction systematically across all included
studies; doing so would be a worthwhile contribution in itself, since
the two architectures carry different uncertainty guarantees and are
frequently reported under the same label.

A canonical fusion pipeline combines ultrasonic, RGB-D, IMU, and GNSS
streams through an extended Kalman filter (EKF) or particle filter,
recursively estimating the user's state vector $\mathbf{x}_t$ and
covariance $\mathbf{P}_t$
\cite{mahammad2026fusion,silva2025navigation}. Silva and
Wimalaratne~\cite{silva2025navigation} instantiate this in a wearable
prototype with a dual-stage perception pipeline: two inclined
ultrasonic sensors on a waist belt are fused homogeneously by EKF to
suppress acoustic noise, and the high-frequency sonar loop then acts
as a trigger: the camera remains idle until an object enters a
critical distance threshold, at which point a single frame is captured
for recognition. This asynchronous coupling is an elegant response to
the battery and thermal limits of wearable hardware. Their field
evaluation with 10 participants (2 blind, 3 with low vision, 5
blindfolded) reports 98\% frontal detection but only
67\% for drop-offs, a safety-critical shortfall arising
from wide-beam ultrasonic geometry at the acute angles of negative
obstacles.

Distance estimation illustrates the same accessibility trade-off in
miniature. Active depth sensing (LiDAR, structured light) yields
centimeter accuracy but is hardware-exclusive, while monocular methods
run on any device.
Salman Shah \emph{et al.}~\cite{salmanshah2026vision} pair a depth
camera with TensorRT-optimized YOLOv5n on a Jetson Nano, announcing
only objects within a distance threshold. The dominant software approach recovers metric
distance from perspective projection, estimating range $Z$ from an
assumed real-world object height $H$, the focal length in pixels
$f_{\mathrm{px}}$, and the bounding-box height $h_{\mathrm{px}}$:
\begin{equation}
Z = \frac{H \cdot f_{\mathrm{px}}}{h_{\mathrm{px}}}.
\label{eq:monocular}
\end{equation}
Variants either regress distance jointly with detection
\cite{vajgl2022distyolo} or fuse a detector with a monocular
depth-estimation network \cite{jaimin2023estimating}. Because
Eq.~\eqref{eq:monocular} recovers range from an inverse dependence on
$h_{\mathrm{px}}$, it inherits the detector's bounding-box error and
degrades with distance as the box shrinks toward the pixel noise
floor. We are not aware of a published monocular error curve for this
estimator within the assistive corpus; the closest available
calibration is the stereo case, where Strbac \emph{et
al.}~\cite{strbac2020yolo} measured roughly 4\% error at 10~m rising
above 50\% at 30~m. That figure is indicative rather than transferable,
because disparity and box height are different quantities, but the
mechanism driving the degradation is the same, and the absence of a
monocular equivalent is itself a gap worth closing. The pragmatic resolution now visible in
the literature is opportunistic: use platform depth APIs where
present and fall back to Eq.~\eqref{eq:monocular} otherwise, which
preserves universal deployability while exploiting better sensing when
the device offers it \cite{das2020realtime}.

Radar--vision fusion projects millimeter-wave detections into the
image plane as dynamic regions of interest, focusing the vision model
and reducing computation \cite{zhang2019radar}. SLAM supplies
continuous localization and incremental mapping for indoor--outdoor
guidance \cite{xu2025travel,bai2017cloud,real2019navigation}: laser
SLAM using LiDAR for dense tracking, and visual SLAM driven by
monocular or stereo cameras for textureless interiors
\cite{xu2025travel}. Deep semantic SLAM extends geometric mapping with
detector-assigned risk categories, fusing streams through Kalman
filters or CNN--LSTM structures, with paths generated by
graph-search algorithms ($A^*$, Dijkstra) or reinforcement learning
policies (DQN, PPO) \cite{xu2025travel}.

Cross-modal audio-visual architectures extend perception beyond the
camera's field of view. Segment Beyond View fuses first-person video
with spatial binaural audio through an Omni2Ego teacher--student
distillation, segmenting vehicles that approach entirely outside the
user's line of sight \cite{wu2024segment}. This directly formalizes the
insight of Section~\ref{sec:era1}: hearing already functions as a
second spatial channel for BLV travelers, and a system that consumes
audio as sensor input rather than merely occupying it as an output
channel respects that fact. Wearable-scale realizations now appear in
high-impact venues: the cross-modal glasses-plus-smartphone device of
Gao \emph{et al.}~\cite{gao2025wearable} performs depth-aided video
compression on a custom FPGA before transmission, reporting
100\% collision avoidance, $<$320~ms latency,
and $\approx$11~h battery life, although the perfect
avoidance rate warrants scrutiny as a controlled-condition result.

\subsubsection{Improving Prediction Accuracy}
\label{ssec:accuracy}
Alarood and Ibrahim~\cite{alarood2026factors} argue that effectiveness
in this domain is determined less by raw model capability than by the
interaction of safety, responsiveness, reliability, and usability. This
framing positions the accuracy work below as necessary but not
sufficient. Within that framing, four strategies recur.

\paragraph{Depth-Augmented Detection}
Okolo \emph{et al.}~\cite{okolo2025smart} pair YOLOv8 on a
Raspberry~Pi with an RGB camera and an ultrasonic transducer to
generate monocular depth estimates, using Eq.~\eqref{eq:ultrasonic}
for proximity. Across nine obstacle classes the
system reports 91.70\% average accuracy, with
92.40\% for a person, 92.00\% for a car, and
94.00\% for a chair at 0.2~s. The source table
also lists a 0.001~s detection time for the person and car classes;
this is implausible as an end-to-end latency on Raspberry Pi hardware
and the source does not document what that figure measures, so it is
omitted here rather than reproduced.

\paragraph{Hybrid Global--Local Feature Extraction}
To resolve localization ambiguity in unstructured urban scenes,
Alarood \emph{et al.}~\cite{alarood2025enhanced} combine a Bilateral
Vision Transformer with a Convolution Kernel Neural Network
(BViT-CKNN). Preprocessing applies an edge-preserving bilateral filter
that replaces each pixel intensity with a weighted average combining a
spatial kernel $SK$ and a range kernel $RK$,
\begin{equation}
\begin{split}
S_{\mathrm{filtered}}(p_i) = \frac{1}{Z} \sum_{q \in \Omega}
&SK\!\left(\lVert q - p_i \rVert\right)\\
&\times RK\!\left(\lVert I(q) - I(p_i) \rVert\right) I(q),
\end{split}
\label{eq:bilateral}
\end{equation}
where $Z$ is the normalization constant and $\Omega$ the neighborhood.
The ViT branch partitions the filtered frame into non-overlapping
$16 \times 16$ patches, flattens them, and projects them through a
learnable matrix to dense embeddings augmented with positional
encodings; the CKNN branch extracts local features through multi-scale
kernels; and global dependencies are resolved by IoU-based multi-head
self-attention \cite{alarood2025enhanced}. The combination reports
precision 0.93, recall 0.91, F1
0.92, and mean absolute error reduced to 3.16\%.
The unaddressed question is latency: transformers are markedly heavier
than the MobileNet architectures the authors critique, and for a
walking user a high-precision model that lags by half a second is more
dangerous than a slightly less precise one that does not.

\paragraph{Multi-Task Embedded Architectures}
Xu \emph{et al.}~\cite{xu2025travel} deploy a multi-task network on an
NVIDIA Jetson Nano with an Intel RealSense D435i depth camera. The
backbone uses an Efficient Depthwise Activation Block with depthwise
separable convolutions alongside a Multi-Scale Context Attention unit;
within it, a spatial reconstruction block applies group normalization
and a learnable scaling parameter $\gamma$ to compute channel
selection weights,
\begin{equation}
W = \sigma\!\left(W_{\gamma} \cdot X_{\mathrm{out}}\right),
\label{eq:srb}
\end{equation}
separating valid semantic streams from redundant channels, and a
progressive channel attention block computes inter-channel weights by
scaled dot-product attention,
\begin{equation}
A = \mathrm{softmax}\!\left(\frac{Q K^{\top}}{\sqrt{C}}\right).
\label{eq:cpab}
\end{equation}
A dual-stream fusion module unfolds features into one-dimensional
sequences across four scanning directions and applies a
state-space recursion,
\begin{equation}
h_t = A h_{t-1} + B x_t, \qquad y_t = C h_t + D x_t,
\label{eq:mamba}
\end{equation}
with an adaptive context-aware head combining grouped convolutions,
deformable convolutions, and squeeze-and-excitation blocks to achieve
75.7\% mAP for obstacle detection and 97.9\% mask
mAP for walkable-road segmentation simultaneously
\cite{xu2025travel}.

\paragraph{Surface Transition Detection}
Sudden changes in walking surface are among the most critical hazards
for BLV travelers, yet most systems address only discrete obstacles.
Mungdee \emph{et al.}~\cite{mungdee2025lowcost} target this gap with a
Pathway Surface Transition Point Detection (PSTPD) method on a
low-cost cane using two RGB cameras (one horizontal for obstacles,
one tilted 25$^{\circ}$ downward for surfaces) running a YOLOv8n model
over eight surface classes including Braille blocks, crosswalks,
holes, puddles, and uneven flooring. Given two adjacent surface
regions with pixel areas $A_1$ and $A_2$, the transition point is
estimated by inverse-area weighting,
\begin{equation}
w_1 = \frac{A_2}{A_1 + A_2}, \qquad
w_2 = \frac{A_1}{A_1 + A_2},
\label{eq:pstpd}
\end{equation}
yielding a weighted center $(C_x, C_y)$ that is mapped to a
pre-calibrated depth reference by linear interpolation and offset by
the fixed camera-to-user distance. Severity is assigned as
$\mathrm{Level}(S_{i,j}) = \max\!\left(\mathrm{Level}(c_i),
\mathrm{Level}(c_j)\right)$, prioritizing surface hazards over
obstacle alerts in the feedback queue. Reported distance error is
0.3~cm for obstacles and 4.22~cm for surface
transitions, with mAP 0.70 and 0.92
respectively \cite{mungdee2025lowcost}.

\subsubsection{Internet Independence via Local Hosting}
\label{ssec:offline}
Localized edge hosting addresses a critical vulnerability: continuous
connectivity fails precisely where users are most exposed: basements,
underground transit, and areas with poor cellular coverage
\cite{kuriakose2023deepnavi,choudri2025vision,datta2024enhanced}.
Executing inference on native hardware eliminates transmission
latency, preserves privacy, and reduces power draw by avoiding
high-bandwidth wireless telemetry
\cite{kuriakose2023deepnavi,choudri2025vision}.

The foundational approach compiles single-stage networks into
optimized binaries for constrained processors. Kuriakose \emph{et
al.}~\cite{kuriakose2021towards} convert a YOLOv5m model pre-trained
on MS COCO \cite{lin2014coco} into a quantized TensorFlow Lite binary
that processes frames directly on embedded CPUs and GPUs. DeepNAVI
goes further architecturally, deploying EfficientDet-Lite4 with an
EfficientNet-Lite4 backbone, a bidirectional feature pyramid network,
and shared box predictors under focal loss, running entirely onboard
\cite{kuriakose2023deepnavi}.

Geometric reasoning is likewise performed locally. Detected bounding
boxes are mapped to screen-space regions to estimate left, center, or
right position, while distance is computed offline by single-camera
geometry, using the Rule of 57 to derive it from the ratio of an object's angular size
to a full circle, checked against a pre-calibrated reference height
and the phone's internal position sensors
\cite{kuriakose2023deepnavi,kuriakose2021towards}. Motion tracking uses
a dual-mode single Gaussian model over image grids with Gaussian and
median pre-filtering, compensating for user-induced background shift
via Kanade--Lucas--Tomasi tracking and homography estimation at
5.8~ms per frame \cite{yi2013detection,kuriakose2023deepnavi}.
Scene recognition runs in parallel on an EfficientNet-Lite4 baseline
with a confidence threshold of 0.7 to suppress false
positives offline \cite{kuriakose2023deepnavi}. Application-level
autonomy is completed by on-device voice command handling and local
haptic loops that scale vibration intensity with proximity, with user
preferences held in an encrypted local store rather than a remote
service \cite{choudri2025vision}.

Between full cloud dependence and full offline operation lies the
dual-mode architecture. Datta \emph{et al.}~\cite{datta2024enhanced}
separate system logic into parallel online and offline paths monitored
by handshaking protocols and hardcoded termination windows: in stable
network conditions the client streams frames to a server running a
full YOLOv4 backbone; if the connection breaks or a response misses
its window, the phone drops the server loop and activates a local
OpenCV-based recognition module that sustains core safety
tasks, namely face tracking and staircase-edge recognition, until
connectivity returns. Baig \emph{et al.}~\cite{baig2026aibased}
likewise keep collision warning on a simple distance sensor and buzzer,
leaving a large vision-language model on a hat-mounted Raspberry~Pi~4
for contextual description and one-click enrollment of new people and
objects (90\% recognition in good lighting).

A recent commercial-hardware study offers a
complementary data point: comparing a local open-vocabulary
object-detection pipeline on smart glasses against a cloud
vision-language model (Meta AI on Ray-Ban Meta glasses) for an
object-reaching task, average completion time fell from 126~s to
53~s with the local pipeline ($p<0.001$), and transcript analysis
attributed part of the gap to spatial hallucination in the cloud
model; in one trial it directed a participant to search the wrong
side of the room for roughly four minutes \cite{singh2026reaching}.
The result is from a single study with five participants and does not
generalize on its own, but it illustrates concretely why local and
cloud inference are increasingly treated as serving different needs
rather than competing on one metric.

\subsubsection{Model Optimization for Edge Deployment}
\label{ssec:opt}
Migrating detectors onto resource-constrained terminals requires
restructuring baseline architectures to minimize parameter count and
FLOPs while holding accuracy within acceptable bounds
\cite{zhou2022mobileyolo,ju2025lsyolo,cai2021yolobile,pedoeem2018yololite}.
Table~\ref{tab:optimization} synthesizes the techniques recurring
across the corpus; we organize them into five groups.

\begin{table}[!t]
\tbl{Synthesis of edge-optimization techniques across the corpus.
Reported gains are \emph{not} directly comparable across rows: they
derive from different baselines, datasets, and hardware, and several
sources do not state whether a latency figure covers inference alone
or the end-to-end pipeline.}
{\begin{tabular}{>{\raggedright\arraybackslash}p{2.48cm}>{\raggedright\arraybackslash}p{3.60cm}>{\raggedright\arraybackslash}p{3.44cm}>{\raggedright\arraybackslash}p{2.96cm}}
\toprule
Technique & Mechanism & Reported effect (example) & Sources\\
\midrule
Backbone substitution & MobileNetV2/V3, ShuffleNetv2+Ghost replace darknet & Large parameter reduction at modest mAP cost & \cite{zhou2022mobileyolo,zeng2023lightweight,ma2024apple,yang2025mfd}\\
\midrule
Depthwise separable conv. & Factorize spatial and channel filtering & FLOP reduction in backbone and head & \cite{zhou2022mobileyolo,li2024improved}\\
\midrule
Structural re-param.\ (CSS) & Multi-branch training, single-conv inference & Higher throughput, lower memory-access cost & \cite{zheng2023stackyolo}\\
\midrule
Partial conv.\ / ASPPF / VoVGSCSP & Restrict or serialize neck computation & $>$116 FPS reported & \cite{yang2025mfd,zheng2023stackyolo,li2024improved}\\
\midrule
Lightweight attention (ECA/ECSA/EMA) & Re-weight salient channels and scales, Eq.~\eqref{eq:eca} & Recovers accuracy lost to compression & \cite{zhou2022mobileyolo,yu2024lpyolo,li2024improved,zhang2023mbab}\\
\midrule
Channel pruning (BN-$\gamma$) & Remove low-scale channels and neighbors & 64.2\% storage cut, $>2\times$ speed, $-4.2\%$ acc. & \cite{cheng2024yolov8}\\
\midrule
Block-punched pruning & Regular sparse blocks, GPU-friendly shape & Real-time mobile detection & \cite{cai2021yolobile}\\
\midrule
PTQ (int8 / fp16) & TensorRT, TFLite, or NCNN conversion & $>3\times$ acceleration; int8 slower without HW support & \cite{wang2023qyolo,zeng2023lightweight,cheng2024yolov8,leong2026tensorflow}\\
\midrule
Knowledge distillation (AWKD) & Teacher--student, Eq.~\eqref{eq:awkd} & Compression with bounded accuracy loss & \cite{yang2025mfd}\\
\midrule
FPGA acceleration & Pipelining, loop unrolling, dynamic pruning & Lower latency and power than CPU/GPU & \cite{vaithianathan2024fpga,gao2025wearable}\\
\midrule
Compiler co-design & GPU/CPU execution-graph partitioning & Parallel execution, reduced latency & \cite{cai2021yolobile,chen2015glimpse}\\
\midrule
Metaheuristic tuning (TSO-STO) & Global search over synaptic weights & Avoids local minima without gradient overhead & \cite{bathool2026tuned}\\
\bottomrule
\end{tabular}}
\label{tab:optimization}
\end{table}

\paragraph{Backbone Light-Weighting}
The dominant strategy substitutes heavy backbones with efficient
alternatives: MobileNetV2/V3 replacements for darknet layers
\cite{zhou2022mobileyolo,zeng2023lightweight}, ShuffleNetv2 and Ghost
hybrids \cite{ma2024apple}, and systematic replacement of standard
convolutions with depthwise separable convolutions (DSC), which
factorize spatial and channel filtering to cut floating-point
operations in backbone and head alike
\cite{zhou2022mobileyolo,li2024improved}. A MobileNet-MCA backbone on
a YOLOv7-tiny baseline models feature relationships across vertical,
horizontal, and channel branches via permutation
\cite{yang2025mfd}; alternatively the native
focus layer can be replaced with a standard down-sampling block plus
MobileNetV3 bneck structures and structured neck pruning
\cite{zeng2023lightweight}. Hardware-friendly structural units such as
Cross, Split, and Shuffle (CSS) blocks enrich gradient combination
while reducing memory-access cost, fusing multi-branch structures into
single convolutions at inference through re-parameterization
\cite{zheng2023stackyolo}. In LP-YOLO, purpose-built lightweight units, namely a basic LP unit
and an LP down-sampling module, pair DSC with channel shuffle to
promote group information flow \cite{yu2024lpyolo}.

\paragraph{Neck Optimization and Attention}
Multi-scale fusion layers are streamlined with partial convolution,
which restricts computation to a subset of input channels
\cite{yang2025mfd}; with adaptive spatial pyramid pooling that
replaces parallel pooling with an attention-driven serial
configuration learning feature weights from context
\cite{zheng2023stackyolo}; and with VoVGSCSP bottlenecks whose hybrid
convolutions, batch normalization, and channel shuffle sustain
reported throughput above 116~FPS \cite{li2024improved}. To
recover accuracy lost to compression, lightweight attention modules
re-weight salient features. Efficient Channel Attention computes an
adaptive kernel size
\begin{equation}
k \;=\; \left| \frac{\log_2 C}{\gamma} + \frac{b}{\gamma}
\right|_{\mathrm{odd}},
\label{eq:eca}
\end{equation}
where $C$ is the number of channels, $\gamma$ and $b$ are the slope and
offset of the linear mapping from $\log_2 C$ to the kernel size
(typically $\gamma = 2$ and $b = 1$), and $|\cdot|_{\mathrm{odd}}$
denotes the nearest odd integer. The resulting kernel enables
cross-channel interaction without dimensionality reduction \cite{zhou2022mobileyolo}. ECSA unifies channel and spatial
re-calibration within a single layer, avoiding sequential attention
delay \cite{yu2024lpyolo}, and Efficient Multiscale Attention splits
input into sub-features to aggregate cross-spatial detail for
severely occluded small targets
\cite{li2024improved,zhang2023mbab}.

\paragraph{Pruning, Quantization, and Distillation}
Sparsity training on the batch-normalization scale factor $\gamma$
provides the evaluation criterion for isolating and removing redundant
channels together with their adjacent convolution structures
\cite{cheng2024yolov8}. For finer granularity, block-punched pruning
divides a layer's weight tensor into equal spatial blocks and removes
identical weight positions across parallel filter and channel
dimensions, preserving a regular matrix shape that avoids thread
divergence and memory read conflicts on mobile GPUs
\cite{cai2021yolobile}. Post-training quantization via TensorRT,
TensorFlow Lite, or NCNN converts 32-bit floating-point parameters to
int8 or half-precision fp16
\cite{yang2025mfd,ikram2025detr,cheng2024yolov8,wang2023qyolo,leong2026tensorflow,li2019fullyquantized}.
Combined BN-pruning and int8 quantization reduced YOLOv8n storage by
64.2\% and more than doubled inference speed at a
4.2\% accuracy cost \cite{cheng2024yolov8}, while Q-YOLO's
training-free pipeline reports over $3\times$ acceleration at near
full precision \cite{wang2023qyolo}. One cautionary result deserves
emphasis: Zeng \emph{et al.}~\cite{zeng2023lightweight} found
their int8 model \emph{slower} than fp16 on a smartphone lacking
dedicated 8-bit instructions, a reminder that software optimization
cannot outrun
hardware that does not support it, and that any survey-level
recommendation must be conditioned on the accelerator present.

Adaptive weighted knowledge distillation
transfers capacity from teacher to student through a combined
objective
\begin{equation}
\mathcal{L}_{\mathrm{AWKD}} \;=\; \alpha \,
\mathrm{CE}_{\mathrm{weighted}} \;+\; \beta \,
\mathrm{MSE}_{\mathrm{box}},
\label{eq:awkd}
\end{equation}
with cross-entropy penalties shifted by adaptive category weights
\cite{yang2025mfd}.

\paragraph{Hardware Acceleration and Compiler Co-Design}
FPGA deployments exploit memory pipelining, dynamic threshold pruning,
and loop unrolling to maximize instruction-level parallelism,
outperforming CPU and GPU baselines in both latency and power
\cite{vaithianathan2024fpga,gao2025wearable}. Spatial pyramid pooling
with dynamic depth adapts processing paths at runtime to input feature
complexity through dynamic routing layers \cite{ju2025lsyolo}.
Compiler-level co-design partitions the execution graph, routing heavy
convolutions to the mobile GPU while assigning lightweight parallel
branches and non-convolutional tasks to the CPU
\cite{cai2021yolobile}, and this can be supplemented by on-device
caching modules that absorb transmission latency without interrupting
the notification loop \cite{chen2015glimpse}. End-to-end pipelines
demonstrate the composite result: a MobileNetV3-backboned, pruned,
fp16-quantized detector at 26.5~FPS on a standard
smartphone CPU \cite{zeng2023lightweight}, and a TFLite MobileNetV2-SSD
at 85~ms per frame on a Raspberry~Pi~4B
\cite{leong2026tensorflow}.

\paragraph{Measured Versus Published Latency}
A methodological caution applies to every figure in
Table~\ref{tab:optimization}. The bulk of the compression literature
reports latency on desktop GPUs, yet the deployment target here is a
mid-range phone, so any assistive system claiming real-time
operation should report latency measured on consumer hardware rather
than inherited from a benchmark.

\paragraph{Metaheuristic Weight Optimization}
Beyond structural re-engineering and pruning, global weight
optimization can be enhanced by metaheuristic search that avoids local
minima without incurring gradient-calculation overhead. A hybrid
strategy combining Tuna Swarm Optimization (TSO) and Siberian Tiger
Optimization (STO) exploits complementary properties: TSO maintains
broad global exploration across high-dimensional decision spaces,
while STO contributes local exploitation and rapid convergence.
Coupling the two tunes the synaptic weights of a lightweight feature
extractor (an improved SqueezeNet) for recognition and the parameter
configuration of a deep residual network for distance estimation
\cite{bathool2026tuned}. This is currently a single-study report: no
second source in the corpus corroborates the TSO--STO approach.

\begin{table}[!t]
\tbl{Benchmarks for the core vision--language tasks underlying
assistive perception.}
{\setlength{\tabcolsep}{4pt}
\begin{tabular}{>{\raggedright\arraybackslash}p{1.75cm}>{\raggedright\arraybackslash}p{2.0cm}>{\raggedright\arraybackslash}p{1.6cm}>{\raggedright\arraybackslash}p{2.1cm}}
\toprule
Task & Key dataset & Metric & Assistive utility\\
\midrule
Panoptic segmentation & ADE20K \cite{zhou2019semantic} & Panoptic Quality & Obstacle detection and environmental mapping\\
\midrule
Depth estimation & NYU Depth v2 \cite{silberman2012indoor} & RMSE & Indoor spatial awareness, drop-off avoidance\\
\midrule
Object detection & MS COCO \cite{lin2014coco} & mAP & General detection; lacks BLV-relevant classes\\
\midrule
Image captioning & VizWiz-Captions \cite{gurari2018vizwiz} & BLEU-1, CIDEr & Scene narration from user-captured views\\
\midrule
Visual question answering & VizWiz-VQA \cite{gurari2018vizwiz,bigham2010vizwiz} & VQA accuracy; hallucination rate & Interactive, task-specific queries\\
\bottomrule
\end{tabular}}
\label{tab:datasets}
\end{table}

\subsection{Cross-Corpus System Comparison}
\label{sec:systems}

Table~\ref{tab:systems} compares representative complete systems from
all four eras along the dimensions extracted in
Section~\ref{sec:method}. Three caveats govern its reading. First,
reported accuracies derive from different datasets and object
vocabularies and are not mutually comparable. Second, latency figures
depend on undisclosed or heterogeneous hardware, and several are
reported without specifying whether they measure inference alone or
the end-to-end pipeline. Third, and most consequentially for
RQ3, the BLV-user-study column is sparsely populated: the
overwhelming majority of systems are evaluated on benchmark metrics
alone.

\Needspace{14\baselineskip}
{\setlength{\tabcolsep}{4pt}
\setlength{\LTleft}{0pt}\setlength{\LTright}{0pt}\tabledim=0pt\tablefont\renewcommand\toprule{\hline}\renewcommand\midrule{\hline}\renewcommand\bottomrule{\hline}
\refstepcounter{table}\edef\LTtabno{\thetable}\label{tab:systems}
\begin{longtable}{>{\raggedright\arraybackslash}p{\dimexpr 0.145\textwidth-2\tabcolsep\relax}>{\centering\arraybackslash}p{\dimexpr 0.080\textwidth-2\tabcolsep\relax}>{\centering\arraybackslash}p{\dimexpr 0.070\textwidth-2\tabcolsep\relax}>{\raggedright\arraybackslash}p{\dimexpr 0.135\textwidth-2\tabcolsep\relax}>{\raggedright\arraybackslash}p{\dimexpr 0.140\textwidth-2\tabcolsep\relax}>{\raggedright\arraybackslash}p{\dimexpr 0.150\textwidth-2\tabcolsep\relax}>{\raggedright\arraybackslash}p{\dimexpr 0.165\textwidth-2\tabcolsep\relax}>{\raggedright\arraybackslash}p{\dimexpr 0.115\textwidth-2\tabcolsep\relax}}
\multicolumn{8}{p{\dimexpr\textwidth-2\tabcolsep\relax}}{\tablecaptionfont{\tablenumfont Table\nobreakspace\LTtabno.}\hskip7pt\ignorespaces Representative assistive navigation systems across the four
eras. US = ultrasonic; TTS = text-to-speech; RPi = Raspberry Pi;
``BLV study'' = evaluation with blind or low-vision participants;
n.r.\ = no such evaluation reported in the source consulted. The
column was assigned by a full-text keyword search of each source for
participant- and user-study-related language. Five of the 22 systems
listed report testing with confirmed BLV participants, and one further
system reports a single-participant pilot of unconfirmed BLV status.}\\
\toprule
System & Era & Year & Sensing & Model & Platform / feedback & Reported performance & BLV study\\
\midrule
\endfirsthead
\toprule
System & Era & Year & Sensing & Model & Platform / feedback & Reported performance & BLV study\\
\midrule
\endhead
\bottomrule
\endfoot
Das \emph{et al.} \cite{das2020realtime} & II & 2020 & US + PIR cane & thresholding & cane; haptic+audio & 400 cm range, velocity sensing & n.r.\\
\midrule
Kasabekar \emph{et al.} \cite{kasabekar2024smart} & II & 2024 & 3$\times$US + GPS & thresholding & modular wearable; haptic+\allowbreak audio+\allowbreak SMS & 10--300 cm range & n.r.\\
\midrule
Chintu \emph{et al.} \cite{chintu2026realtime} & II & 2026 & US + GPS (IoT) & thresholding & wearable; vibration+audio & real-time proximity + tracking & n.r.\\
\midrule
Pham \emph{et al.} \cite{pham2016realtime} & II/III & 2016 & Kinect RGB-D & RANSAC geometry & PC; audio & high wall/floor-object acc. & n.r.\\
\midrule
Bai \emph{et al.} \cite{bai2017cloud} & III & 2017 & stereo helmet cam & cloud SLAM + CNN + OCR & cloud; audio & 1000+ classes & n.r.\\
\midrule
Bai \emph{et al.} \cite{bai2019wearable} & III/IV & 2019 & RGB-D + IMU glasses & lightweight CNN & smartphone; audio & indoor + outdoor nav. & 20 participants (20 BLV)\\
\midrule
Joshi \emph{et al.} \cite{joshi2020efficient} & III & 2020 & camera + US & YOLOv3 & standalone; audio & 95.19\% (custom set) & 36 participants (20 BLV)\\
\midrule
Mukhiddinov \& Cho \cite{mukhiddinov2021smart} & III & 2021 & smart glasses & enhancement + salient detection & glasses; TTS+tactile & low-light robustness & n.r.\\
\midrule
Suman \emph{et al.} \cite{suman2022vision} & III & 2022 & cane cam + US shoe & SSD + RNN & cane+shoe; audio+alert & 95.06\% indoor; 4.82 s latency & n.r.\\
\midrule
Ashiq \emph{et al.} \cite{ashiq2022cnn} & III & 2022 & phone cam + GPS & MobileNet & smartphone; audio & 83.3\% & 4 participants (1 BLV)\\
\midrule
Kuriakose \emph{et al.} \cite{kuriakose2023deepnavi} & IV & 2023 & phone camera & EfficientDet-Lite & smartphone only; audio & 20 custom classes, real-time & 1 participant (pilot)\\
\midrule
Bougheloum \emph{et al.} \cite{bougheloum2024outdoor} & III & 2024 & wearable camera & YOLOv5 + GTTS & earphones; audio & outperforms R-CNN baseline & n.r.\\
\midrule
Datta \emph{et al.} \cite{datta2024enhanced} & IV & 2024 & phone camera & YOLOv4 dual-mode & phone+server; audio & 70--80\% distance est. & n.r.\\
\midrule
Vijetha \& Geetha \cite{vijetha2024obstackle} & IV & 2024 & phone camera & YOLOv3-tiny + zoning & smartphone; TTS & 92.3\% mAP daytime & n.r.\\
\midrule
Okolo \emph{et al.} \cite{okolo2025smart} & IV & 2025 & RPi cam + US + moisture & YOLOv8 & RPi + phone app; voice & 91.7\% avg. & n.r.\\
\midrule
Gao \emph{et al.} \cite{gao2025wearable} & IV & 2025 & glasses video + depth & cross-modal learning & FPGA + phone; audio+tactile & 100\% avoidance, $<$320 ms, 11 h & 12 participants (12 BLV)\\
\midrule
Mungdee \emph{et al.} \cite{mungdee2025lowcost} & IV & 2025 & 2$\times$cam + US cane & YOLOv8n + PSTPD & smart cane; alerts & mAP 0.92 surface, 0.70 obstacle & n.r.\\
\midrule
Silva \& Wimalaratne \cite{silva2025navigation} & IV & 2025 & US + vision + IMU + GNSS & EKF / ES-EKF fusion & belt; tactile+audio & 98\% frontal, 67\% drop-off & 10 participants (5 BLV)\\
\midrule
Wu \emph{et al.} \cite{wu2024segment} & IV & 2024 & ego cam + binaural audio & Omni2Ego distillation & (AR HMD target) & +11.4\% OOV mIoU vs.\ SOTA & n.r.\\
\midrule
Leong \emph{et al.} \cite{leong2026tensorflow} & IV & 2026 & RPi cam + US & MobileNetV2-SSD (TFLite) & RPi 4B; buzzer+TTS & 92.5\%, 85 ms/frame & n.r.\\
\midrule
Baskar \emph{et al.} \cite{baskar2026enhancing} & IV & 2026 & phone camera & YOLOv8n + ego-motion fusion & edge; audio & 79.9\% mAP@0.5, 21 FPS & n.r.\\
\midrule
Prakash \emph{et al.} \cite{prakash2026aipowered} & IV & 2026 & webcam + US + GSM & CNN detection + face rec. & RPi; audio+SMS & 96\% clear, 95\% face rec. & n.r.\\
\end{longtable}}

\subsection{Datasets, Benchmarks, and Evaluation Tasks}
\label{sec:datasets}

Benchmark fragmentation is among the field's most serious structural
problems. General-purpose datasets such as MS COCO \cite{lin2014coco}
dominate training pipelines \cite{bougheloum2024outdoor,kuriakose2021towards},
yet their class vocabularies omit the hazards that matter most to BLV
travelers, such as curbs, surface transitions, drop-offs, and
head-level obstructions. In response, many groups build small custom datasets: 20
navigation-relevant classes \cite{kuriakose2023deepnavi}, 7{,}600
labeled obstacle images \cite{bastidas2025computer}, a
wheelchair-perspective curb/sidewalk/gutter set \cite{kao2024edge},
eight pathway-surface classes \cite{mungdee2025lowcost}, and ten
outdoor categories fused with COCO \cite{bougheloum2024outdoor}.
The consequence across the corpus is that reported
accuracies (Table~\ref{tab:systems}) are mutually incommensurable, and
no public leaderboard exists for BLV navigation perception.

Table~\ref{tab:datasets} summarizes the principal benchmarks used to
evaluate the vision--language capabilities that increasingly underpin
assistive systems. The VizWiz project remains the foundational
BLV-authentic resource: introduced in 2010 as a near-real-time
question-answering service \cite{bigham2010vizwiz} and later released
as multi-task datasets captured directly by blind users
\cite{gurari2018vizwiz}, it forces algorithms to cope with the
low-quality, poorly framed, and blurred inputs that characterize real
use. Its limitation for this survey's purposes is scope: it targets
visual question answering rather than navigation.

Two structural gaps follow. First, generalist multimodal models
underperform on accessibility-specific tasks even when they excel at
general-domain questions, exhibiting elevated hallucination rates and
weakness on optical Braille recognition, recognition of assistive
objects such as white canes and tactile guide paths, and multilingual
input \cite{karamolegkou2025evaluating}. Second, most benchmarks rely
on static images, whereas real-world navigation is a temporal process;
egocentric video question-answering work is beginning to address this
gap \cite{song2024video}, but the field still needs broader
video-based evaluation that models temporal reasoning and continuous
obstacle avoidance.

\subsection{User-Centred Evaluation and Adoption}
\label{sec:users}

Applying the simplicity constraint of Section~\ref{sec:era1} across
the corpus yields this survey's most sobering finding: technical
evaluation has largely displaced human evaluation. Belli \emph{et
al.}~\cite{belli2026navigation} quantify the structural irony
precisely: 92.7\% of solutions are described as human-centered
wearable systems, yet only 39\% involved BLV individuals during
initial design, and 50\% require dedicated user training as a
consequence. Our own corpus reproduces the pattern. Of the 22
representative systems compared in Table~\ref{tab:systems}, five
report evaluation involving confirmed BLV participants: Joshi
\emph{et al.}~\cite{joshi2020efficient} with 36 testers (20 visually
impaired, 16 blindfolded), Ashiq \emph{et al.}~\cite{ashiq2022cnn} with
four participants of whom one was visually impaired, Silva and
Wimalaratne~\cite{silva2025navigation} with ten (two blind, three with
low vision, five blindfolded), Gao \emph{et
al.}~\cite{gao2025wearable} with twelve blind and visually impaired
volunteers evaluated over seven months of real-world use (the longest
evaluation window in the table), and Bai \emph{et
al.}~\cite{bai2019wearable} with twenty visually impaired participants
(ten totally blind, ten partially sighted) recruited under a protocol
approved by a disabled-persons' federation. That cohort is the largest
in the table and, notably, the only one of the five with no
blindfolded-proxy substitution at all. A further system, Kuriakose
\emph{et al.}'s DeepNAVI~\cite{kuriakose2023deepnavi}, reports a
single-participant pilot whose BLV status the source does not state.
Two observations follow. First, this set still leans on blindfolded
sighted proxies for roughly a quarter of participants (21 of 82
across the five confirmed studies), a substitution that the
disability-studies literature warns does not reproduce the adaptive
strategies of people with lived experience
\cite{bennett2018interdependence}. Second, the median study size is
twelve participants, still too small to generalize across a population
as heterogeneous as this one, though Gao \emph{et al.}'s seven-month
deployment and Bai \emph{et al.}'s ethics-board-approved, all-BLV
cohort are notable exceptions to the field's otherwise brief and
proxy-heavy evaluation practice.

This inversion matters because the documented failure modes of
assistive technology are overwhelmingly human factors: cognitive
overload from continuous alerts
\cite{chavan2026aipowered,real2019navigation,chanana2017assistive},
steep learning curves among older users \cite{chavan2026aipowered},
excessive cost \cite{nikanfar2025survey,chanana2017assistive}, and the
social awkwardness of conspicuous hardware
\cite{chavan2026aipowered,real2019navigation}. Where user studies do
exist, they surface problems no benchmark would: Silva and
Wimalaratne~\cite{silva2025navigation} report that testers experienced
confusion when multiple waist-belt vibration motors fired
simultaneously in crowded spaces, meaning that a system designed to
reduce cognitive load reintroduced it through an unprioritized haptic
channel.

Design responses visible in the corpus include salient-object
filtering to suppress redundant announcements
\cite{mukhiddinov2021smart}; audio trimming, object counting
(``three persons'' rather than three separate announcements), and
category limiting to the three-to-five most relevant classes per frame
\cite{joshi2020efficient}; separation of channels so that spatial
distance is conveyed haptically while audio is reserved for semantic
labels \cite{silva2025navigation}; and DeepNAVI's deliberate
restriction to a single familiar device and modality
\cite{kuriakose2023deepnavi}. Bone-conduction audio is a further
response, leaving the ear canal open to ambient sound while still
delivering feedback \cite{kuriakose2023deepnavi}.
Belli \emph{et al.}'s prescription of co-design with BLV users from the
earliest architectural phase \cite{belli2026navigation} should be
treated not as a recommendation but as a methodological requirement.

\section{Discussion}

\subsection{Ethics, Privacy, and Equity}
\label{sec:ethics}

Always-on wearable cameras record bystanders who have not consented,
and cloud-tethered pipelines transmit that imagery off-device
(Section~\ref{sec:era3}); edge-native processing
(Section~\ref{sec:era4}) is therefore a privacy architecture as much
as a latency one. Continuous passive capture can record home
addresses, banking details, and medical documents alongside the faces,
voices, and activities of passers-by. The asymmetry is documented:
Akter \emph{et al.}~\cite{akter2022shared} surveyed 128 visually
impaired wearers and 136 sighted bystanders and found a genuine
conflict of interest: BLV users prioritize dynamic behavioral
attributes such as expressions, gestures, and activities for social
navigation, while bystanders are least willing to share precisely
those attributes, particularly when captured by smart glasses whose
extended field of view exceeds natural human sight lines. Lee
\emph{et al.}~\cite{lee2020pedestrian} reproduced the asymmetry in an
in-person study pairing blind participants wearing a detection
prototype with sighted passers-by, and observed that bystanders became
\emph{more} reserved after learning what the system could infer. The
picture is further complicated by the fact that BLV individuals are
themselves bystanders to other people's cameras, and cannot visually
detect a recording device in order to avoid it
\cite{zhao2023sighted}. Proposed mitigations center on \emph{privacy
frictions}: physical shutter indicators, audible capture tones, and
localized LED indicators that establish a framework of notice and
consent \cite{ahmed2018bystander,akter2022shared}.

Security engineering remains rare in the corpus. The AAL smart cane of
Tsantikidou \emph{et al.}~\cite{tsantikidou2025aal} is a notable
exception, encrypting LoRa communication with the lightweight CLEFIA
cipher to protect data exchanged with mobile devices and cloud
services while preserving real-time performance. Chavan \emph{et
al.}~\cite{chavan2026aipowered} advocate federated learning so that
model updates occur without harvesting raw individual telemetry, and
Choudri \emph{et al.}~\cite{choudri2025vision} keep user preferences in
an encrypted local store.

A second risk is algorithmic misrepresentation. Akter \emph{et
al.}~\cite{akter2022shared} identify this as the \emph{shared} ethical
concern uniting users and bystanders: imperfect models may misclassify
a bystander's expression, activity, or intent, and the harm ranges
from ordinary recognition errors to the questionable application of
inference to subjective social constructs. A third risk concerns
training-data provenance. Sighted engineers frequently generate data
through disability simulation, such as blindfolds and blur filters, but
simulated data does not reproduce the behavioral patterns and
adaptive strategies of people with years of lived experience, and
simulation can foster pity and paternalistic design that reduces users
to a set of deficits \cite{bennett2018interdependence}.

Equity concerns compound the picture. The global BLV population is
concentrated among older adults and in low-resource settings
\cite{who2019vision}, while many proposed systems presume recent
smartphones, stable connectivity, or bespoke hardware; Nikanfar
\emph{et al.}~\cite{nikanfar2025survey} note that LiDAR and AR, the
most capable spatial technologies, are precisely the least affordable,
creating a socioeconomic barrier that contradicts the inclusive
premise of assistive technology. Low-cost designs
\cite{ikram2025obstacle,vennila2025smartiot,mungdee2025lowcost,leong2026tensorflow}
address affordability directly, but systematic cost reporting is
absent from most studies. Finally, safety liability is unexamined: no
study in the corpus specifies failure-mode guarantees for a system
whose false negative may cause physical injury, and regulatory
conformance is equally unexamined: no study reports how its design
satisfies statutory obligations such as the Americans with
Disabilities Act, the European Accessibility Act (in force since June
2025), the accessibility standards that increasingly anchor such
regulation \cite{w3c2023wcag}, or the data-protection constraints that
frameworks like the GDPR place on continuously captured egocentric
video.

A framing shift is worth registering. Accessibility research has
historically pursued absolute independence through technology, but
Bennett \emph{et al.}~\cite{bennett2018interdependence} argue for
\emph{interdependence}, a concept drawn from disability studies and
disability justice, as a complementary frame: disabled people live within
collaborative networks, and technology should support ability-diverse
cooperation rather than isolating the user inside automated
assistance. Read through that lens, two design elements this survey
treats as engineering details become substantive commitments: the
graceful-handoff behavior discussed in Section~\ref{sec:challenges}
is not a fallback for model failure but a design goal, and the
co-constructed perception observed between travelers and non-human
guides \cite{due2023guidedog} is a template rather than an
inefficiency.

\subsection{Open Challenges and Future Directions}
\label{sec:challenges}

\subsubsection{Seamless Indoor--Outdoor Continuity}
No surveyed system transitions gracefully between GPS-denied interiors
and satellite-guided exteriors \cite{belli2026navigation}; hybrid
SLAM-plus-infrastructure approaches
\cite{xu2025travel,real2019navigation} remain partial, and
infrastructure-free indoor localization at cane-relevant accuracy is
unsolved. Negative obstacles are a specific instance of the same
problem: drop-off detection at 67\%
\cite{silva2025navigation} is not a research curiosity but a
life-safety gap, and closing it likely requires UWB or infrared ToF
sensing rather than wide-beam ultrasound.

\subsubsection{Benchmark and Protocol Standardization}
The field needs a public, BLV-authentic navigation benchmark with an
ego-viewpoint, hazard-centric classes, and standardized latency and
hardware reporting, so that the incommensurable accuracy claims of
Table~\ref{tab:systems} can become comparable science
\cite{gurari2018vizwiz}. Belli \emph{et al.}~\cite{belli2026navigation}
identify the enabling condition: only about one paper in seven
currently publishes open data.

\subsubsection{Cognitive-Load-Aware Interfaces}
Perception quality has outrun interface quality. Salience filtering
\cite{mukhiddinov2021smart}, audio compression \cite{joshi2020efficient},
and modality budgeting \cite{kuriakose2023deepnavi} point the way, but
principled, user-calibrated information-rate control (what to say,
when, and through which channel) remains open, as does rigorous
measurement of cognitive load during real navigation
\cite{real2019navigation,chavan2026aipowered}. One direction the surveyed
systems leave open is \emph{magnitude-proportional haptics}: rather
than announcing an object's class, a device might vibrate in
proportion to the approaching object's size or threat level (a light
pulse for a bollard, a strong one for a vehicle), conveying urgency
through a pre-attentive channel without consuming the auditory one.
Elements of this exist in isolation. Several systems already modulate
pulse frequency by proximity
\cite{silva2025navigation,kasabekar2024smart}, and Mungdee \emph{et
al.}~\cite{mungdee2025lowcost} compute an explicit per-hazard severity
level; what is absent is a system that maps severity rather than
distance onto vibration magnitude, and an evaluation of whether users
can decode such an encoding reliably under load. We propose this as an
open question rather than a finding.

\subsubsection{Multimodal Foundation Models}
Vision--language models promise conversational scene understanding
beyond fixed class vocabularies, answering ``where is the door
handle?'' rather than announcing ``door: 87\%'', and early
conversational architectures already gesture in this direction
\cite{chavan2026aipowered}. Recent head-mounted evaluations sharpen
the picture: open-vocabulary prompting on smart glasses allows a user
to specify an arbitrary target and receive stereo directional audio
cues, and a direct comparison against a cloud vision-language model on
an object-reaching task found the local pipeline faster and less
prone to the spatial hallucination that produced incorrect directions
from the cloud model \cite{singh2026reaching}.

Reliability, not capability, is the binding constraint. A hallucinated
response in a consumer application is an inconvenience; a hallucinated
response about a medication dose, a crosswalk signal, or an obstacle
can cause physical harm. The design protocols proposed for
multimodal visual assistants follow from this asymmetry: neutral
factual communication conveying only verifiable visual facts without
speculation; adaptive communication that establishes user preferences
early; goal-oriented filtering of irrelevant background;
\emph{transparent uncertainty handling} that explicitly states when an
image is too dark, blurred, or ambiguous to answer; and \emph{graceful
handoff} to human assistance when a task exceeds the model's
reliability limits \cite{gonzalezpenuela2026visionpal}. The remaining
open problems are the ones this survey traces across every era:
latency and energy on wearable hardware, and
privacy for continuously streamed egocentric video.

\subsubsection{Alternative and Emerging Interfaces}
Robotic guide dogs (Section~\ref{sec:era1}) transplant a proven
interaction model onto a programmable platform and are the nearest of
the alternatives to maturity, with a dedicated systematic review
already available \cite{hong2022guidedogrobot} and empirical
comparison against biological guide dogs under way
\cite{due2023guidedog}. Their open problems are the ones this survey
identifies throughout: cost, transport practicality, and the fact
that a distributed, co-constructed model of perception has not yet
been formalized into design requirements.

\subsubsection{Convergence with Distracted-Pedestrian Safety}
A thread running quietly through the recent corpus is that BLV
navigation and distracted-pedestrian safety are the same perception
problem arising from two different causes of reduced situational
awareness. Smartphone distraction while walking measurably degrades
awareness and balance, raising the incidence of collisions, trips, and
falls \cite{stavrinos2018distracted,pelicioni2023mobile}, and several
systems in this survey are explicitly framed as serving both groups:
jaywalking detection with event-triggered alerts
\cite{mostafi2022jaywalking}, smartphone-sensor hazard anticipation
\cite{islam2021realtime}, and quantized on-device detection motivated by
both populations \cite{wang2023qyolo}.

The convergence is technically real, since the detector, the latency
budget, and the compression pipeline are shared, but it should not be
overstated. The two populations differ in expertise, in the baseline
mobility skill they bring, in how much attention a system may
legitimately demand, and in the consequence of a false negative.
Whether the shared substrate justifies a shared product, or whether it
merely creates a larger market for a system that serves neither group
well, is an open question the field has not examined empirically.

\subsubsection{From Prototype to Product}
Closing the 60\% prototype gap \cite{belli2026navigation} requires
co-design, longitudinal field deployment in genuine operational
contexts rather than university corridors, cost transparency, and
open data sharing. These are organizational practices, not
algorithms.

\section{Conclusion}
\label{sec:conclusion}

This survey traced assistive navigation for BLV individuals across
four technological eras, from the white cane's tactile immediacy to
edge-native deep learning systems that fuse camera, ultrasonic,
inertial, and satellite sensing on hardware that fits in a pocket. Two
threads bind the genealogy together. First, a capability thread: each
era answered its predecessor's defining deficit: ETAs added the
sensing the cane lacked, deep learning added the semantics ETAs
lacked, and edge optimization removed the connectivity dependence that
made cloud intelligence unsafe at walking pace. Second, a constraint
thread: the simplicity and cognitive-load requirements that explain
the white cane's endurance remain binding on every successor, and the
systems that honor them by filtering salience, budgeting modalities,
separating haptic from semantic channels, and running silently on
familiar devices are the ones most likely to escape the laboratory.

The evidence assembled here suggests the field's bottleneck has
migrated. It is no longer detection accuracy, which is high and still
improving, nor inference latency, which the compression techniques of
Section~\ref{ssec:opt} have brought within the envelope of consumer
hardware. It is the translation of perception into information a
walking person can act on without being overwhelmed, and the
willingness to validate that translation with the people who will use
it. The technical trajectory now points toward multimodal,
conversational, on-device intelligence; whether that trajectory
produces adopted tools rather than accumulated prototypes will depend
on the field treating BLV users as co-designers, standardizing its
benchmarks, and engineering for privacy and affordability as
first-class requirements rather than afterthoughts.

\bibliographystyle{tfnlm}
\bibliography{references}

\end{document}